\documentclass[aps,pra,amsmath,amssymb,floatfix,twocolumn,amsmath,superscriptaddress,twocolumn,nofootinbib,tighten,letterpaper]{revtex4-2}

\usepackage{mathptmx}

\usepackage{graphicx}% Include figure files
\usepackage{dcolumn}% Align table columns on decimal point
\usepackage{bm}% bold math

\usepackage{mwe} 

\usepackage{upgreek}
\usepackage{braket}
\usepackage{xcolor}

\usepackage[colorlinks,linkcolor=blue,citecolor=blue,urlcolor=blue]{hyperref}
\usepackage{hyperref}

\renewcommand\vec[1]{\ensuremath\boldsymbol{#1}}

\begin{document}

\title{Ordinary lattice defects as probes of topology}

\author{Aiden J.\ Mains}
\thanks{These authors contributed equally to this work.}
\affiliation{Department of Physics, Lehigh University, Bethlehem, Pennsylvania 18015, USA}

\author{Jia-Xin Zhong}
\thanks{These authors contributed equally to this work.\\
Contact author: Jiaxin.Zhong@psu.edu}
\affiliation{Graduate Program in Acoustics, The Pennsylvania State University, University Park, PA 16802, USA}

\author{Yun Jing}\thanks{Contact author: yqj5201@psu.edu}
\affiliation{Graduate Program in Acoustics, The Pennsylvania State University, University Park, PA 16802, USA}

\author{Bitan Roy}\thanks{Contact author: bitan.roy@lehigh.edu}
\affiliation{Department of Physics, Lehigh University, Bethlehem, Pennsylvania 18015, USA}

\begin{abstract}
In addition to topological lattice defects such as dislocations and disclinations, crystals are also accompanied by unavoidable ordinary defects, devoid of any non-trivial geometry or topology, among which vacancies, Schottky defects, substitutions, interstitials, and Frenkel pairs are the most common. In this work, we demonstrate that these ubiquitous ordinary lattice defects, although geometrically trivial, can nonetheless serve as universal probes of the non-trivial topology of electronic Bloch bands, and any change in the local topological environment in an otherwise normal insulator in terms of mid-gap bound states in their vicinity. We theoretically establish these generic findings by implementing a minimal model Hamiltonian describing time-reversal symmetry breaking topological and normal insulators on a square lattice, fostering such point defects. The defect-bound mid-gap modes are also shown to be robust against sufficiently weak random point-like charge impurities. Furthermore, we showcase experimental observation of such bound states by embedding ordinary crystal defects in two-dimensional acoustic Chern lattices, where precision-controlled hopping amplitudes are implemented via active meta-atoms and Green’s-function-based spectroscopy is used to reconstruct spectra and eigenstates. Our combined theory-experiment study establishes ordinary lattice defects as probes of topology that should be germane in crystals of any symmetry and dimension, raising the possibility of arresting localized Majorana modes near such defects in the bulk of topological superconductors and to emulate ordinary-defect-engineered topological devices. 
\end{abstract}

\maketitle

\section{Introduction}

\begin{figure*}[t!]
    \centering
    \includegraphics[width=1.00\linewidth]{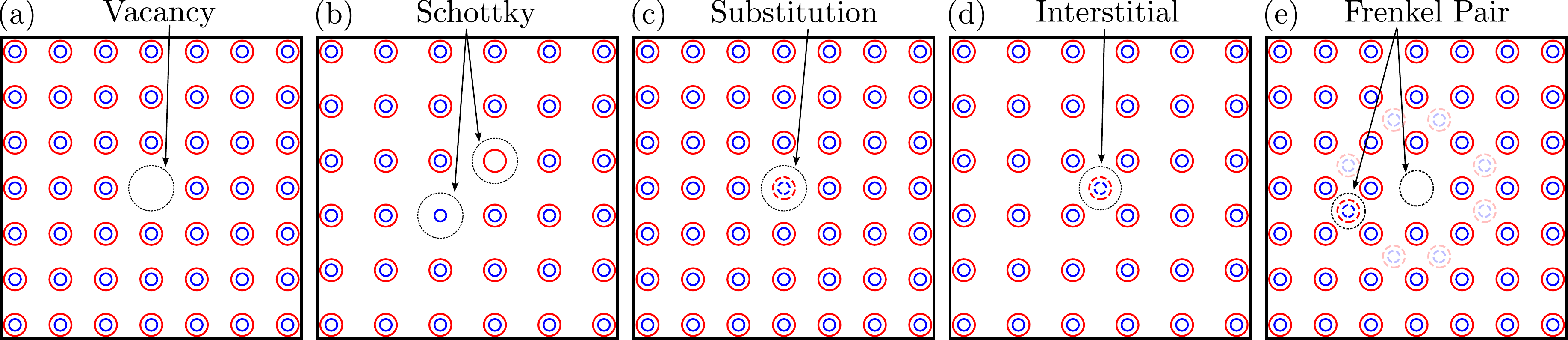}
    \caption{
        Schematic representations of (a) a vacancy, (b) a Schottky defect, (c) a substitution, (d) an interstitial, and (e) a Frenkel pair, where the defect cores are encircled by black dashed closed loops. The solid red and blue circles at each site of the underlying square lattice represent two orbitals with opposite parity eigenvalues ($+$ and $-$) of the Qi-Wu-Zhang model. To create a vacancy defect we eliminate both the orbitals from the same site, whereas to create a Schottky defect we eliminate complementary orbitals from two different sites of the square lattice, whose spatial separation is arbitrary. A substitution defect is generated by setting the value of the on-site orbital staggered potential $m_0$ at a given site in a topologically different parameter regime of the Qi-Wu-Zhang model (represented by the dashed red and blue circles) than the background one (solid red and blue circles). An interstitial defect is built by placing a site at an irregular point within the square lattice with a different on-site orbital staggered potential value in comparison to the background one. A Frenkel pair is a superposition of a vacancy and an interstitial. With a Frenkel pair defect, we average the local density of states of the resulting in-gap bound states (when they exist) over eight positions (marked by one clear and seven translucent dashed red and blue circles) of the constituting interstitial for a fixed vacancy position in our numerical simulations (but not in acoustic lattice-based measurements).     
    }
    \label{fig:1}
\end{figure*}

Defects are ubiquitous in crystals and can broadly be classified as topological or ordinary. Dislocation and disclination are the prominent examples of topological crystal defects that are characterized by their geometric properties. For example, a dislocation is identified by its Burgers vector, while a disclination in a two-dimensional crystal is represented by the corresponding Frank angle. On the other hand, crystals also harbor defects that are devoid of any non-trivial geometric property. Examples of such defects encompass vacancies, Schottky defects, substitutions, interstitial, and Frenkel pairs~\cite{chaikinlubensky, kleinert}. See Fig.~\ref{fig:1} for schematic representations of such ordinary defects on a square lattice.

Over the past decade, numerous examples have endorsed geometrically non-trivial crystal defects, namely dislocations~\cite{dislocation:1, dislocation:2, dislocation:3, dislocation:4, dislocation:5, dislocation:6, dislocation:7, dislocation:8, dislocation:9, dislocation:10, dislocation:11, dislocation:12, dislocation:13, dislocation:14, dislocation:15, dislocation:16, dislocation:17, dislocation:18, dislocation:19, dislocation:20, dislocation:21, dislocation:22} and disclinations~\cite{disclination:1, disclination:2, disclination:3, disclination:4, disclination:5}, as useful tools to recognize and distinguish gapped and nodal topological phases of matter in terms of robust isolated modes, localized around such defect cores that are guaranteed by rigorous mathematical quantization rules and thus are protected against sufficiently weak disorder. Jurisdiction of such crystal defects extends over the landscape of Hermitian (first-order and higher-order), non-Hermitian, and Floquet topological phases, which have also received substantial support from various experiments in quantum materials and meta crystals in classical systems. Furthermore, an array of dislocations, known as grain boundaries can also be instrumental in identifying various topological phases by featuring a mini-band of gapless modes residing along such a ridge, irrespective of its characteristic angle (small or large)~\cite{GB:1, GB:2, GB:3, GB:4, GB:5}. The mini-band of gapless grain boundary modes is also well-separated from the bulk states.

In this work we show, both theoretically and experimentally,  that ordinary lattice defects, despite their geometric triviality, can likewise serve as powerful probes  of topological phases of matter. For concreteness, we consider the square lattice-regularized two-orbital Qi-Wu-Zhang (QWZ) model for time-reversal symmetry breaking (although it plays no role in any conclusion of ours) insulators that depending on the parameter values can be either topological or normal as a demonstrative example~\cite{model:1}. On such a system we introduce vacancies, Schottky defects, substitutions, interstitials, and Frenkel pairs and numerically search for defect-bound localized mid-gap modes that can probe the underlying topology. To eliminate any signature of the edge modes in the topological parameter regime of the QWZ model, throughout we impose periodic boundary conditions in both directions in our numerical simulations as well as during measurements. We arrive at the following key observation (summarized below) from exact numerical diagonalization and by implementing precision-controlled hopping via active meta-atoms in acoustic crystals, fostering such ordinary lattice defects on square lattices. See Fig.~\ref{fig:exp_setup} for a schematic of the experimental setup. 

%%%%%%%%%%%%%%%%%%%%%%%%%%%%%%%%%%%%%%%%%%%%%%%%%%%%%%%%%%%%%%%%%%%%%%
%%%%%%%%%%%%%%%%%%%%%%%%%%%%%%%%%%%%%%%%%%%%%%%%%%%%%%%%%%%%%%%%%%%%%%
%%%%%%%%%%%%%%%%%%%%%%%%%%%%%%%%%%%%%%%%%%%%%%%%%%%%%%%%%%%%%%%%%%%%%%
%%%%%%%%%%%%%%%%%%%%%%%%%%%%%%%%%%%%%%%%%%%%%%%%%%%%%%%%%%%%%%%%%%%%%%
%%%%%%%%%%%%%%%%%%%%%%%%%%%%%%%%%%%%%%%%%%%%%%%%%%%%%%%%%%%%%%%%%%%%%%
%%%%%%%%%%%%%%%%%%%%%%%%%%%%%%%%%%%%%%%%%%%%%%%%%%%%%%%%%%%%%%%%%%%%%%

\subsection{Summary of main results}

We find that both vacancy and Schottky defect tend to create a boundary within the system and thus feature robust mid-gap bound states near the defects only when the crystal is in the topological parameter regime. On the other hand, the remaining three ordinary defects, namely substitution, interstitial, and Frenkel pair, usually respond to any change in local topological environment irrespective of the nature of the background insulator (topological or trivial), in terms of mid-gap bound states, localized near the defects. Particularly, when a trivial insulator hosts a single site or a few sites with topological parameters or vice versa, the system typically features defect-localized bound states. Therefore, ordinary lattice defects are capable of probing any change in local topological environment. In this context our theoretical or numerical results are summarized in Figs.~\ref{fig:2} (vacancy), \ref{fig:3} (Schottky defect), \ref{fig:4} (substitution), and \ref{fig:5} (interstitial and Frenkel pair) and the experimental validation of the key theoretical findings are shown in Figs.~\ref{fig:exp_vacancy}-\ref{fig:exp_Frenkel}. Altogether, the present theory-experiment collaborative venture establishes ordinary lattice defects as useful tools to probe topological phases of matter and any change in local topological environment in an otherwise normal insulators. Such a set of findings should be germane in topological crystals from any symmetry class and in physical dimension. Finally, we numerically show that the mid-gap modes, localized near the defect are immune to weak random point-like charge impurities (Fig.~\ref{fig:append}).

\subsection{Organization}

The remainder of the manuscript is organized as follows. In the next section (Sec.~\ref{sec:model}), we introduce the Bloch Hamiltonian for the QWZ model, express it as a tight-binding Hamiltonian on a square lattice, and discuss its phase diagram in terms of the appropriate topological invariant (the first Chern number and the Bott index). Section~\ref{sec:defecttheory} is devoted to the theoretical studies of the QWZ model in the presence of various ordinary lattice defects in terms of mid-gap bound states therein. In Sec.~\ref{sec:experiment}, we validate some of the main theoretical findings experimentally on time-reversal symmetry breaking acoustic lattices. In Sec.~\ref{sec:summary}, we summarize the main findings and present discussions on possible future extensions of the current study. In Appendix~\ref{appensec:disorder}, we shows that stability of all the mid-gap defect modes against weak random pointlike charge disorder numerically. The localization of mid-gap modes in the close vicinity of ordinary lattice defects is demonstrated from their inverse participation ration in Appendix~\ref{appensec:IPR}.

%%%%%%%%%%%%%%%%%%%%%%%%%%%%%%%%%%%%%%%%%%%%%%%%%%%%%%%%%%%%%%%%%%%%%%
%%%%%%%%%%%%%%%%%%%%%%%%%%%%%%%%%%%%%%%%%%%%%%%%%%%%%%%%%%%%%%%%%%%%%%
%%%%%%%%%%%%%%%%%%%%%%%%%%%%%%%%%%%%%%%%%%%%%%%%%%%%%%%%%%%%%%%%%%%%%%
\begin{figure*}[t!]
    \centering
    \includegraphics[width=0.95\linewidth]{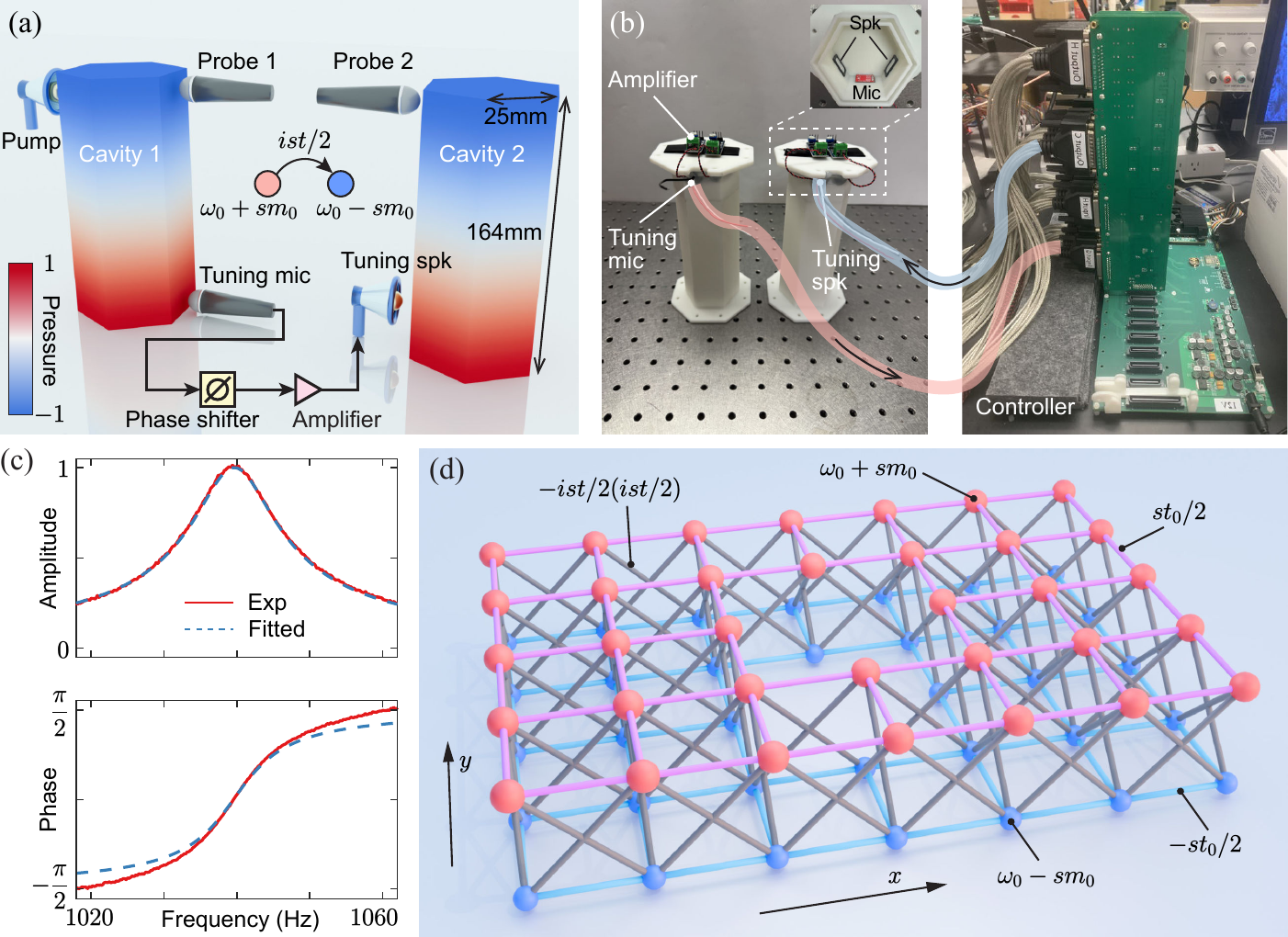}
    \caption{
        (a) A schematic illustration of the unidirectional hopping implementation from Cavity 1 to Cavity 2, representing two sites or orbitals of the Qi-Wu-Zhang model.
        The two cavities should not be interpreted as symmetric and antisymmetric supermodes of two reciprocally coupled resonators.
         The middle inset shows the equivalent tight-binding model with a representative hopping term of $i s t / 2$. 
        A speaker (pump) placed in Cavity 1 provides excitations, while two microphones (probes) record the spectral responses in both cavities. (b) Photograph of the two coupled cavities (left). The amplifier controls the hopping strength, and the controller (right) functions as an integrated phase shifter, as illustrated in (a). Inset in the left panel: interior of the cavity cap, which houses two speakers and one microphone. (c) Experimentally measured (solid lines) and numerically fitted (dashed lines) amplitude (top) and phase (bottom) responses of the cross-power spectral density between the acoustic signals in Cavity 1 and Cavity 2. The target hopping is $i s t / 2 = -0.8 i$, and the fitted value is $-0.82 i$. (d) A schematic of a $7 \times 5$ acoustic lattice with a single vacancy at the center. Red and blue spheres denote two orbitals at each lattice site, while connecting tubes indicate hopping amplitudes. For details see Sec.~\ref{sec:experiment}. Here, $s$ is an overall scale factor. 
    }
    \label{fig:exp_setup}
\end{figure*}
%%%%%%%%%%%%%%%%%%%%%%%%%%%%%%%%%%%%%%%%%%%%%%%%%%%%%%%%%%%%%%%%%%%%%%
%%%%%%%%%%%%%%%%%%%%%%%%%%%%%%%%%%%%%%%%%%%%%%%%%%%%%%%%%%%%%%%%%%%%%%
%%%%%%%%%%%%%%%%%%%%%%%%%%%%%%%%%%%%%%%%%%%%%%%%%%%%%%%%%%%%%%%%%%%%%%

\section{Model and phase diagram}~\label{sec:model}

We begin the discussion with the QWZ model~\cite{model:1}, captured by the Bloch Hamiltonian
\allowdisplaybreaks[4] 
\begin{equation}~\label{eq:QWZBloch}
   H^{\rm Bloch}_{\rm QWZ} = \sum_{\vec{k}} \left( c^\dagger_{+,\vec{k}} \:\: c^\dagger_{-,\vec{k}} \right) \: \left[ \sum^{3}_{j=1}\tau_j d_j(\vec{k}) \right] \: \left( \begin{array}{c}
c_{+,\vec{k}} \\ 
c_{-,\vec{k}} 
\end{array} \right),
\end{equation}
where $c^\dagger_{\tau,\vec{k}}$ and $c_{\tau,\vec{k}}$ are respectively the Fermionic creation and annihilation operators with parity eigenvalues $\tau=\pm$ and momentum $\vec{k}=(k_x,k_y)$, the set of Pauli matrices $\{ \tau_1, \tau_2, \tau_3 \}$ operates on the parity index, and the components of the $\vec{d}$-vector are given by 
\allowdisplaybreaks[4] 
\begin{eqnarray}~\label{eq:dvec}
d_1(\vec{k}) &=& t \sin (k_x a), \:\:
d_2(\vec{k}) = t \sin(k_ya), \nonumber \\
\text {and} \:\: d_3(\vec{k}) &=& m_0 + t_0 \left[ \cos(k_x a) + \cos(k_y a) \right]
\end{eqnarray}
with $a$ as the lattice spacing. The hopping amplitudes between the orbitals with opposite and same parities, living on the nearest-neighbor sites of the underlying square lattice is given by $t$ and $t_0$, respectively, both of which are set to be unity for convenience, without any loss of generality. The on-site staggered density between two orbitals is $m_0$, which determines the topological property of this model that can be captured from the first Chern number of the filled valence band~\cite{model:2, model:3}, for example, given by
\allowdisplaybreaks[4] 
\begin{equation}~\label{eq:chernnumber}
C=-\int_{\rm FBZ} \dfrac{d^{2}{\vec k}}{4\pi} \:\: \big[ \partial_{k_x} \hat{\vec{d}}(\vec{k}) \times \partial_{k_y} \hat{\vec{d}}(\vec{k}) \big] \cdot \hat{\vec{d}}(\vec{k}),
\end{equation}
where $\hat{\vec{d}}(\vec{k})=\vec{d}(\vec{k})/|\vec{d}(\vec{k})|$ and the momentum integral is restricted within the first Brillouin zone (FBZ). Throughout, we neglect particle-hole asymmetry, captured by a term proportional to $\tau_0$ (two-dimensional identity matrix) which does not affect the topology of the system as long as it is an insulator.

%%%%%%%%%%%%%%%%%%%%%%%%%%%%%%%%%%%%%%%%%%%%%%%%%%%%%%%%%%%%%%%%%%%%%%
%%%%%%%%%%%%%%%%%%%%%%%%%%%%%%%%%%%%%%%%%%%%%%%%%%%%%%%%%%%%%%%%%%%%%%
%%%%%%%%%%%%%%%%%%%%%%%%%%%%%%%%%%%%%%%%%%%%%%%%%%%%%%%%%%%%%%%%%%%%%%
\begin{figure*}[t!]
    \centering
    \includegraphics[width=1.00\linewidth]{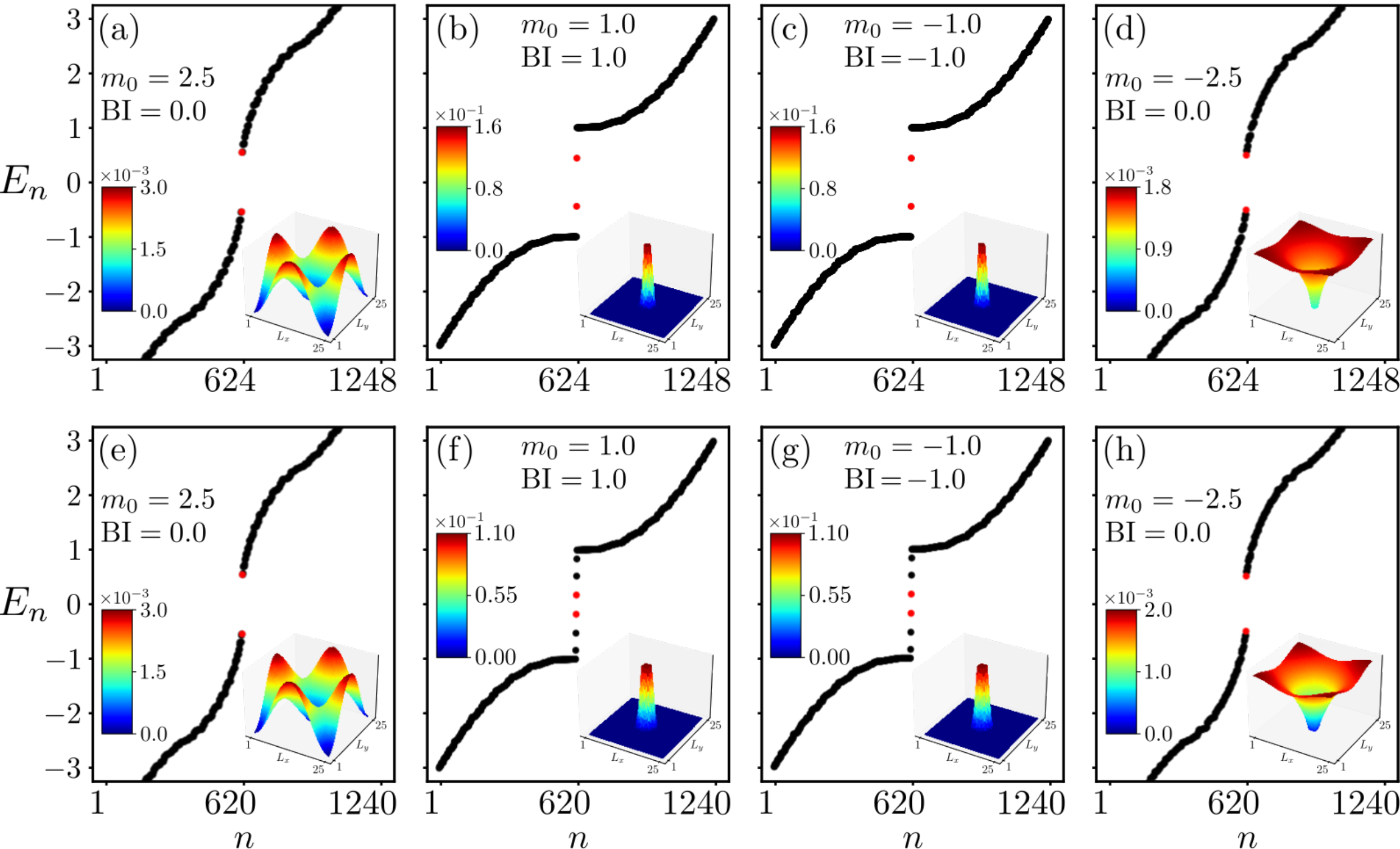}
    \caption{
        Energy eigenvalues ($E_n$) as a function of its index ($n$) in the presence of a single vacancy [(a)-(d)] and five vacancies [(e)-(h)] located at the center of a square lattice with linear dimension $L=25 a$ and periodic boundary conditions in both directions, where $a$ is the lattice spacing. Inset in each panel shows the joint local density of states for two particle-hole symmetric closest-to-zero-energy modes (shown in red). The corresponding value of the mass parameter $m_0$ in the Qi-Wu-Zhang model and the associated Bott index (BI) are quoted in the legend. Therefore, the results are shown for the normal insulator with band minima near the ${\rm M}$ point [(a) and (e)], the ${\rm M}$ phase [(b) and (f)], the $\Gamma$ phase [(c) and (g)], and the normal insulator with band minima near the $\Gamma$ point [(d) and (h)]. See Fig.~\ref{fig:1}(a) for reference and Sec.~\ref{subsec:vacancytheory} for a detailed discussion on the results.   
    }
    \label{fig:2}
\end{figure*}
%%%%%%%%%%%%%%%%%%%%%%%%%%%%%%%%%%%%%%%%%%%%%%%%%%%%%%%%%%%%%%%%%%%%%%
%%%%%%%%%%%%%%%%%%%%%%%%%%%%%%%%%%%%%%%%%%%%%%%%%%%%%%%%%%%%%%%%%%%%%%
%%%%%%%%%%%%%%%%%%%%%%%%%%%%%%%%%%%%%%%%%%%%%%%%%%%%%%%%%%%%%%%%%%%%%%

The QWZ model with the above form supports topological and normal (trivial) insulators within the parameter regime $|m_0/t_0|<2$ and $|m_0/t_0|>2$, respectively. The topological regime, however, fragments into two sectors. Respectively, for $0<m_0/t_0<2$ and $-2<m_0/t_0<0$ the resulting insulators are characterized by $C=+1$ and $C=-1$, featuring the requisite band inversion at the ${\rm M}=(1,1)\pi/a$ and $\Gamma=(0,0)$ points of the FBZ, and are known as the ${\rm M}$ phase and $\Gamma$ phase. Although $C=0$ always in a trivial insulator irrespective of the parameter values $m_0/t_0>2$ and $m_0/t_0<-2$, we note that the minima (maxima) of the non-inverted conduction (valence) band therein occurs at the ${\rm M}$ and $\Gamma$ points, respectively, and are thus characterized by distinct ${\bf K}_{\rm min}$~\cite{model:4}, which plays a crucial role in the forthcoming discussion. Both the topological insulators (${\rm M}$ and $\Gamma$ phases) support topologically robust one-dimensional chiral edge modes, which can be demonstrated from the corresponding tight-binding Hamiltonian on a square lattice ($H^{\rm SL}_{\rm TB}$) obtained by a Fourier transformation of Eq.~\eqref{eq:QWZBloch}. With the given form of the $\vec{d}$-vector in Eq.~\eqref{eq:dvec} and open boundary conditions in both directions $H^{\rm SL}_{\rm TB}$ takes the explicit form 
\allowdisplaybreaks[4] 
\begin{eqnarray}~\label{eq:realspaceTB}
H^{\rm SL}_{\rm TB} &=& \sum_{\vec{r}} \bigg[ m_0 \Psi^{\dagger}_{\vec{r}} \tau_3 \Psi_{\vec{r}}
+ \sum_{j=1,2} \bigg\{ \frac{t}{2i} \Psi^{\dagger}_{\vec{r}} \tau_j \Psi_{\vec{r}+\hat{e}_j} \nonumber \\
&+& \frac{t_0}{2} \Psi^{\dagger}_{\vec{r}} \tau_3 \Psi_{\vec{r}+\hat{e}_j} \bigg\} 
+ H.c. \bigg].
\end{eqnarray}
Here, $\vec{r}=(x,y)$, $\Psi^\top_{\vec{r}}=(c_{\vec{r},+},c_{\vec{r},-})$, $c_{\vec{r},\tau}$ is the fermionic annihilation operator at $\vec{r}$ with parity $\tau=\pm$, and $\hat{e}_j=a \hat{j}$ with $\hat{j}$ as the unit vector along $j=x$ and $y$.

On the other hand, the $\Gamma$ and ${\rm M}$ phases respond distinctly to a dislocation lattice defect, for example, being sensitive to the band inversion momentum ${\bf K}_{\rm inv}=(0,0)$ and $(1,1)\pi/a$, respectively. A dislocation lattice defect is obtained by partially removing a line of atoms and subsequently joining the resulting edges following the Volterra cut-and-paste procedure. Such a defect is characterized by a geometric quantity, the Burgers vector ${\bf b}$ quantifying the missing translation in the close vicinity of the defect core, which when conspires with ${\bf K}_{\rm inv}$ to match a quantization condition ${\bf K}_{\rm inv} \cdot {\bf b}=\pi$ (modulo $2 \pi$), yields topologically robust mid-gap modes in the spectrum that are highly localized around the core of dislocation lattice defect. Naturally, ${\bf K}_{\rm inv} \cdot {\bf b}=0$ always in the $\Gamma$ phase, while ${\bf K}_{\rm inv} \cdot {\bf b}=\pi$ (modulo $2 \pi$) in the ${\rm M}$ phase when ${\bf b}=\pm (2 n+1)a \hat{e}_j$ for $j=x$ or $y$, where $n$ is a positive integer. Therefore, the ${\rm M}$ phase features topologically robust mid-gap modes, while the $\Gamma$ phase is devoid of them~\cite{dislocation:1, dislocation:3, dislocation:5, dislocation:7, dislocation:9, dislocation:10, dislocation:11, dislocation:14, dislocation:18, dislocation:19}. The lattice defects with non-trivial geometry in this way serve as a probe to distinguish topological insulators with different ${\bf K}_{\rm inv}$. In this work, on the other hand, we set out to establish pivotal roles of ordinary lattice defects in identifying topological insulators and any change in the local topological environment in an otherwise normal insulator in terms of bound states in their vicinity, about which more in the next section.

%%%%%%%%%%%%%%%%%%%%%%%%%%%%%%%%%%%%%%%%%%%%%%%%%%%%%%%%%%%%%%%%%%%%%%
%%%%%%%%%%%%%%%%%%%%%%%%%%%%%%%%%%%%%%%%%%%%%%%%%%%%%%%%%%%%%%%%%%%%%%
%%%%%%%%%%%%%%%%%%%%%%%%%%%%%%%%%%%%%%%%%%%%%%%%%%%%%%%%%%%%%%%%%%%%%%
\begin{figure*}[t!]
    \centering
    \includegraphics[width=1.00\linewidth]{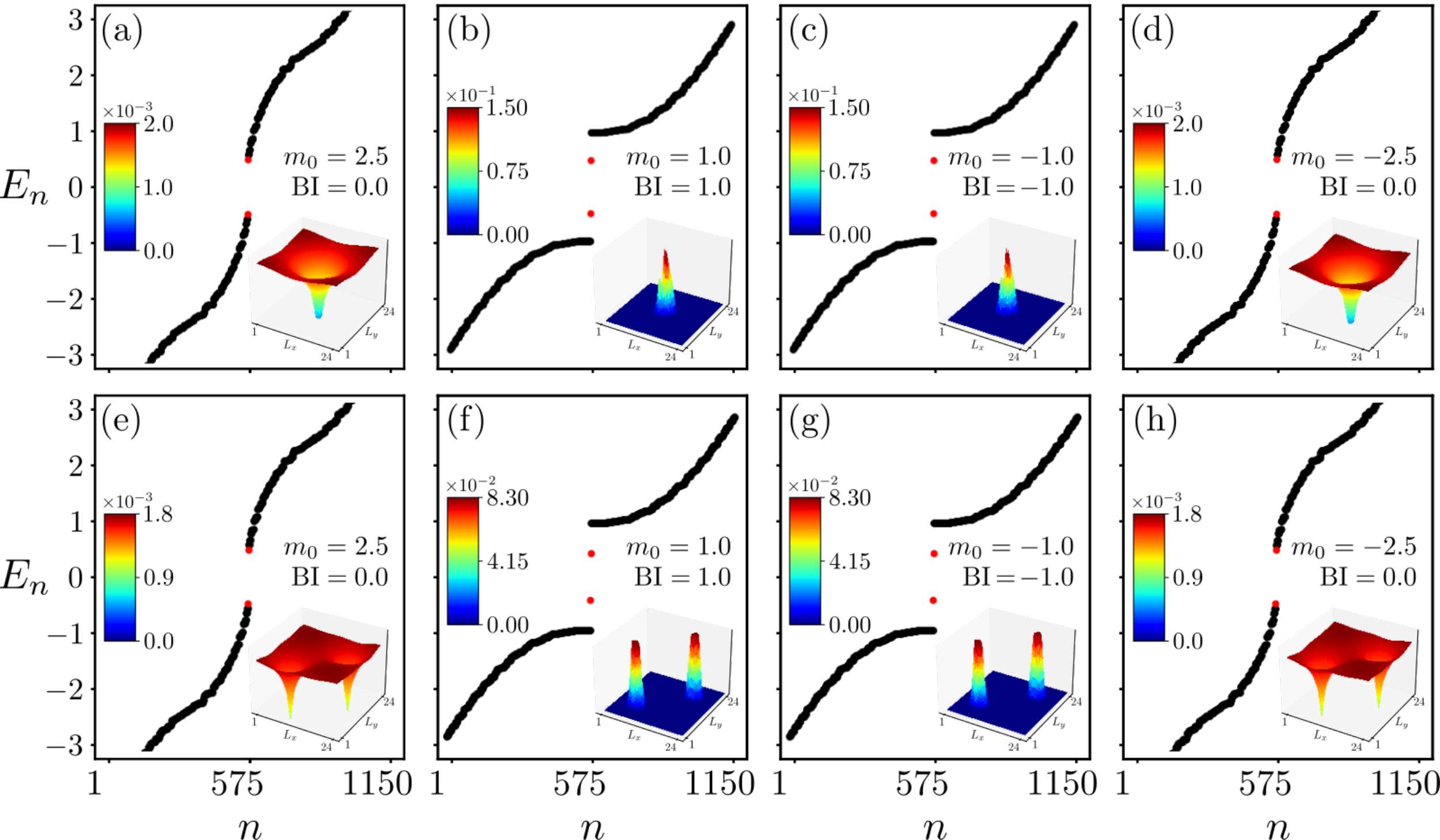}
    \caption{
        Energy eigenvalues ($E_n$) as a function of its index ($n$) in the presence of a single Schottky defect for which the vacancies of opposite-parity orbitals reside at a distance of $\sqrt{2} a$ [(a)-(d)] and $11 \sqrt{2}a$ [(e)-(h)] about the center of a square lattice with linear dimension $L=24 a$ and periodic boundary conditions in both directions, where $a$ is the lattice spacing. Inset in each panel shows the joint local density of states for two particle-hole symmetric closest-to-zero-energy modes (shown in red). The corresponding value of the mass parameter $m_0$ in the Qi-Wu-Zhang model and the associated Bott index (BI) are quoted in the legend. Therefore, the results are shown for the normal insulator with band minima near the ${\rm M}$ point [(a) and (e)], the ${\rm M}$ phase [(b) and (f)], the $\Gamma$ phase [(c) and (g)], and the normal insulator with band minima near the $\Gamma$ point [(d) and (h)]. See Fig.~\ref{fig:1}(b) for reference and Sec.~\ref{subsec:schottkytheory} for a detailed discussion on the results.      
    }
    \label{fig:3}
\end{figure*}
%%%%%%%%%%%%%%%%%%%%%%%%%%%%%%%%%%%%%%%%%%%%%%%%%%%%%%%%%%%%%%%%%%%%%%
%%%%%%%%%%%%%%%%%%%%%%%%%%%%%%%%%%%%%%%%%%%%%%%%%%%%%%%%%%%%%%%%%%%%%%
%%%%%%%%%%%%%%%%%%%%%%%%%%%%%%%%%%%%%%%%%%%%%%%%%%%%%%%%%%%%%%%%%%%%%%

We note that in the presence of disorder or lattice defects, the translational symmetries of the square lattice is broken. Then we cannot rely on the first Chern number to characterize and identify topologically distinct insulators. Fortunately, there exists another topological invariant, namely the Bott index that is equivalent to the first Chern number. However, the Bott index is computed by diagonalizing the real space Hamiltonian $H^{\rm SL}_{\rm TB}$ in the following way. First, we define two diagonal matrices, ${\boldsymbol X}$ and ${\boldsymbol Y}$, with their matrix elements given by $X_{i,j}=x_i \delta_{i,j}$ and $Y_{i,j}=y_i \delta_{i,j}$, respectively, encoding the position $(x_i,y_i)$ of the $i$th site, where $\delta_{i,j}$ is the Kronecker delta symbol. From such diagonal position operators, next we define two periodic unitary matrices ${\boldsymbol U}_X=\exp(2 \pi i {\boldsymbol X}/L)$ and ${\boldsymbol U}_Y=\exp(2 \pi i {\boldsymbol Y}/L)$, where $L$ is the linear dimension of the square lattice in the $x$ and $y$ directions. Next, we define the projector (${\mathcal P}$) onto the filled eigenstates of $H^{\rm SL}_{\rm TB}$, given by $\ket{E}$ with energy $E$ that satisfies $H^{\rm SL}_{\rm TB} \ket{E} = E \ket{E}$, up to the Fermi energy in the half-filled system $E_F=0$ (in the absence of any particle-hole asymmetry), defined as ${\mathcal P}=\sum_{E<E_F} \ket{E}\bra{E}$. The Bott index (BI) is then given by~\cite{model:5} 
\begin{equation}~\label{eq:bottindex}
{\rm BI}= \frac{1}{2 \pi} {\Im} \left( {\rm Tr} \left[ \ln \left( {\bf V}_X {\bf V}_Y {\bf V}^\dagger_X {\bf V}^\dagger_Y \right) \right]\right),
\end{equation}   
in systems with periodic boundary conditions, where ${\bf V}_X= {\boldsymbol I}-{\mathcal P} + {\mathcal P} {\boldsymbol U}_X {\mathcal P}$ and ${\bf V}_Y= {\boldsymbol I}-{\mathcal P} + {\mathcal P} {\boldsymbol U}_Y {\mathcal P}$ with ${\boldsymbol I}$ as the unity matrix. We find that $C \equiv {\rm BI}$ always. Note that during the computation of the Bott index, we always impose periodic boundary conditions in both directions. The additional hopping elements that ensure such boundary conditions are not shown explicitly in Eq.~\eqref{eq:realspaceTB} for brevity.

Finally, note that the square lattice-based tight-binding model $H^{\rm SL}_{\rm TB}$ from Eq.~\eqref{eq:realspaceTB}, stemming from the specific form of the ${\vec d}$-vector from Eq.~\eqref{eq:dvec} in the QWZ model allows hopping processes only along two principle axes between the nearest-neighbor sites residing at a distance of $a$ in the $x$ and $y$ directions. On the other hand, in the presence of interstitial and Frenkel pair lattice defects, a site is introduced at an \emph{irregular point} that does not coincide with any site location of the underlying square lattice [see Figs.~\ref{fig:1}(d) and~\ref{fig:1}(e)]. In order to ensure that such a newly introduced site gets duly coupled with its surrounding ones we generalize $H^{\rm SL}_{\rm TB}$ in the following way. We employ the method of symmetry, introduced in Refs.~\cite{model:6, model:7}, and replace each term appearing in the $\vec{d}$-vector, constituting the Bloch QWZ Hamiltonian [Eq.~\eqref{eq:QWZBloch}], by its symmetry analogous term in the real space, such that both transform identically under all the discrete symmetry operations, namely, the four-fold rotations about the $z$ direction, and the reflections about the $x$ and $y$ axes. Then the resulting generalized tight-binding model takes the explicit form 
\allowdisplaybreaks[4]
\begin{eqnarray}~\label{eq:Method1}
&&H^{\rm SL, gen}_{\rm TB} = \sum_{j \neq k} \exp\left[ 1-\frac{r_{jk}}{r_0}\right] \; \frac{\Theta(R - r_{jk})}{2} \; c^\dagger_j \bigg[\frac{t}{i} \nonumber \\
&\times& \big( \tau_1 \cos \phi_{jk}  
+ \tau_2 \sin \phi_{jk} \big)
+ t_0 \tau_3 \bigg] c_k + m_0 \sum_{j} c^\dagger_j \tau_3 c_j, 
\end{eqnarray}
where $r_{jk}=|\vec{r}_j-\vec{r}_k|$ and $\phi_{jk}$ are the distance and azimuthal angle between the $j$th and $k$th sites, respectively, located at $\vec{r}_j$ and $\vec{r}_k$, and $c^\top_j=\left( c_{+,j}, c_{-,j} \right)$ is a two-component spinor with $c_{\tau,j}$ as the fermion annihilation operator with parity $\tau=\pm$ on the $j$th site. The exponential factor controls the hopping amplitude that decays with the radial distance between two sites and we choose $r_0=a$. The Heaviside step function ($\Theta$) truncates the hopping processes beyond a certain radial distance $R$. To mimic the nearest-neighbor hopping Hamiltonian that keeps all the sites only within a distance of $a$ connected, we choose $R=a$. With these parameterizations, the phase diagram of $H^{\rm SL, gen}_{\rm TB}$ in terms of the Bott index is identical to that for $H^{\rm SL}_{\rm TB}$ from Eq.~\eqref{eq:realspaceTB} and $H^{\rm Bloch}_{\rm QWZ}$ from Eq.~\eqref{eq:QWZBloch}. Variations of the tunable parameters $R$ and $r_0$ only cause nonuniversal shift of the phase boundaries between the the topological and normal insulating phases, without affecting their existence in the global phase diagram. For example, with $R=a$ but $r_0=3 a$, the $\Gamma$ phase with ${\rm BI}=-1$ and the ${\rm M}$ phase with ${\rm BI}=+1$ are realized for $-4<m_0/t_0<0$ and $0<m_0/t_0<4$, respectively.

With all the necessary tools in hand, we now proceed to scrutinize the responses of the QWZ model, featuring Chern and normal insulators in the presence of ordinary lattice defects, mentioned earlier. In parallel, we also compute the Bott index to keep track of the topological invariant of the system in the presence of such defects. 

%%%%%%%%%%%%%%%%%%%%%%%%%%%%%%%%%%%%%%%%%%%%%%%%%%%%%%%%%%%%%%%%%%%%%%
%%%%%%%%%%%%%%%%%%%%%%%%%%%%%%%%%%%%%%%%%%%%%%%%%%%%%%%%%%%%%%%%%%%%%%
%%%%%%%%%%%%%%%%%%%%%%%%%%%%%%%%%%%%%%%%%%%%%%%%%%%%%%%%%%%%%%%%%%%%%%
\begin{figure*}[t!]
    \centering
    \includegraphics[width=1.00\linewidth]{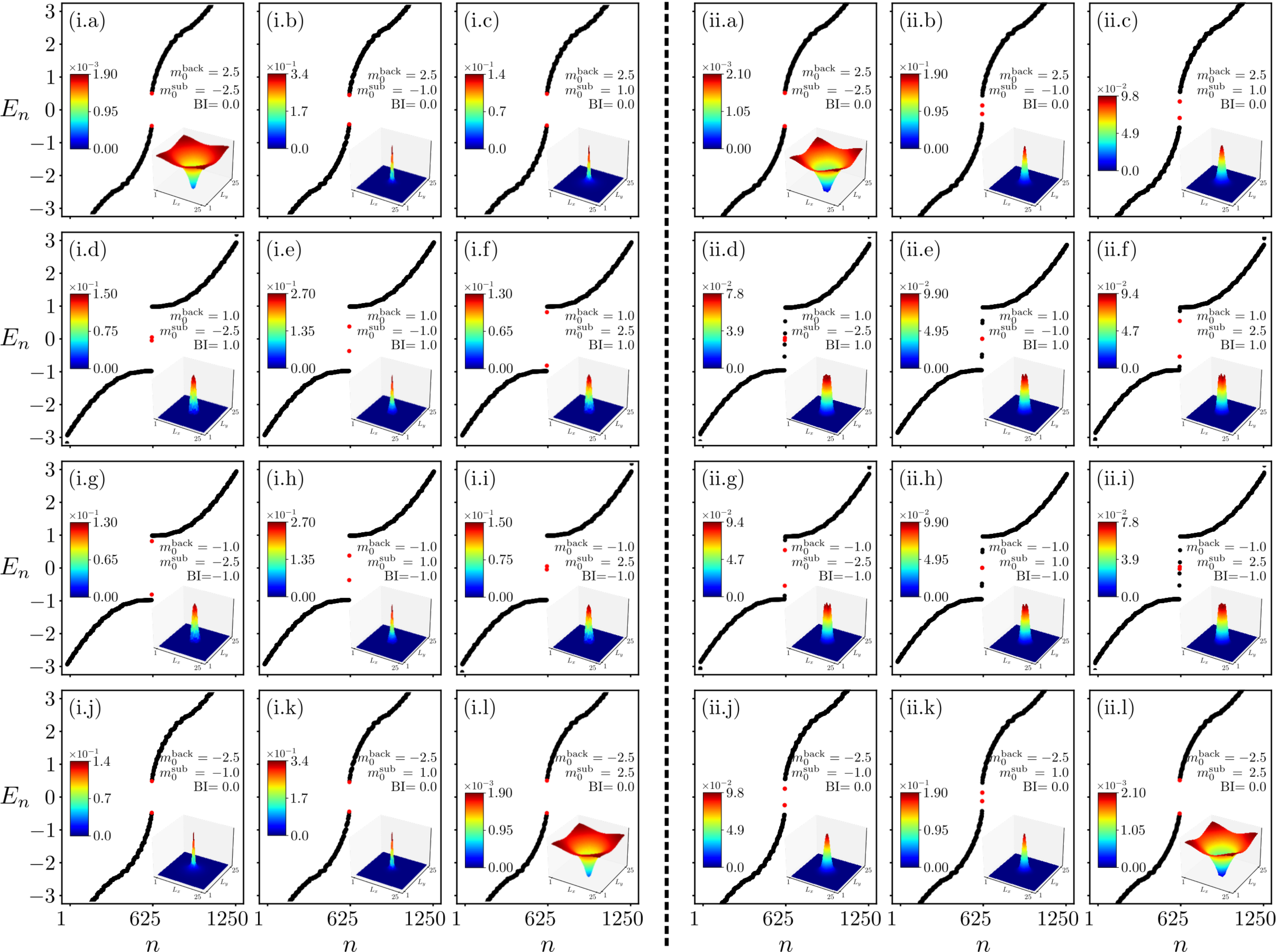}
    \caption{
             Energy eigenvalues ($E_n$) as a function of its index ($n$) in the presence of a single substitution [(i.a)-(i.l)] and five substitutions [(ii.a)-(ii.l)] located at the center of a square lattice with linear dimension $L=25 a$ and periodic boundary conditions in both directions, where $a$ is the lattice spacing. Inset in each panel shows the joint local density of states for two particle-hole symmetric closest-to-zero-energy modes (shown in red). The corresponding value of the background mass parameter $m^{\rm back}_0$ on the parent square lattice sites and substituted mass parameter $m^{\rm sub}_0$ on the substitution defect sites in the Qi-Wu-Zhang model, respectively represented by the solid and dashed circles (red and blue) in Fig.~\ref{fig:1}(c), and the associated Bott index (BI) are quoted in the legend of each panel. See Sec.~\ref{subsec:substitutiontheory} for a detailed discussion on the results.       
    }
    \label{fig:4}
\end{figure*}
%%%%%%%%%%%%%%%%%%%%%%%%%%%%%%%%%%%%%%%%%%%%%%%%%%%%%%%%%%%%%%%%%%%%%%
%%%%%%%%%%%%%%%%%%%%%%%%%%%%%%%%%%%%%%%%%%%%%%%%%%%%%%%%%%%%%%%%%%%%%%
%%%%%%%%%%%%%%%%%%%%%%%%%%%%%%%%%%%%%%%%%%%%%%%%%%%%%%%%%%%%%%%%%%%%%%

\section{Ordinary lattice defects and bound states}~\label{sec:defecttheory}

In this section, we discuss the constructions of the ordinary lattice defects within the framework of the QWZ model in two dimensions and discuss the emergent bound states near such ordinary defects along with their possible topological origins. We devote the next five subsections to cover this journey. In this section we only present and discuss results obtained from exact numerical diagonalizations on square lattices with periodic boundary conditions in both directions.

%%%%%%%%%%%%%%%%%%%%%%%%%%%%%%%%%%%%%%%%%%%%%%%%%%%%%%%%%%%%%%%%%%%%%%
%%%%%%%%%%%%%%%%%%%%%%%%%%%%%%%%%%%%%%%%%%%%%%%%%%%%%%%%%%%%%%%%%%%%%%
%%%%%%%%%%%%%%%%%%%%%%%%%%%%%%%%%%%%%%%%%%%%%%%%%%%%%%%%%%%%%%%%%%%%%%
\begin{figure*}[t!]
    \centering
    \includegraphics[width=1.00\linewidth]{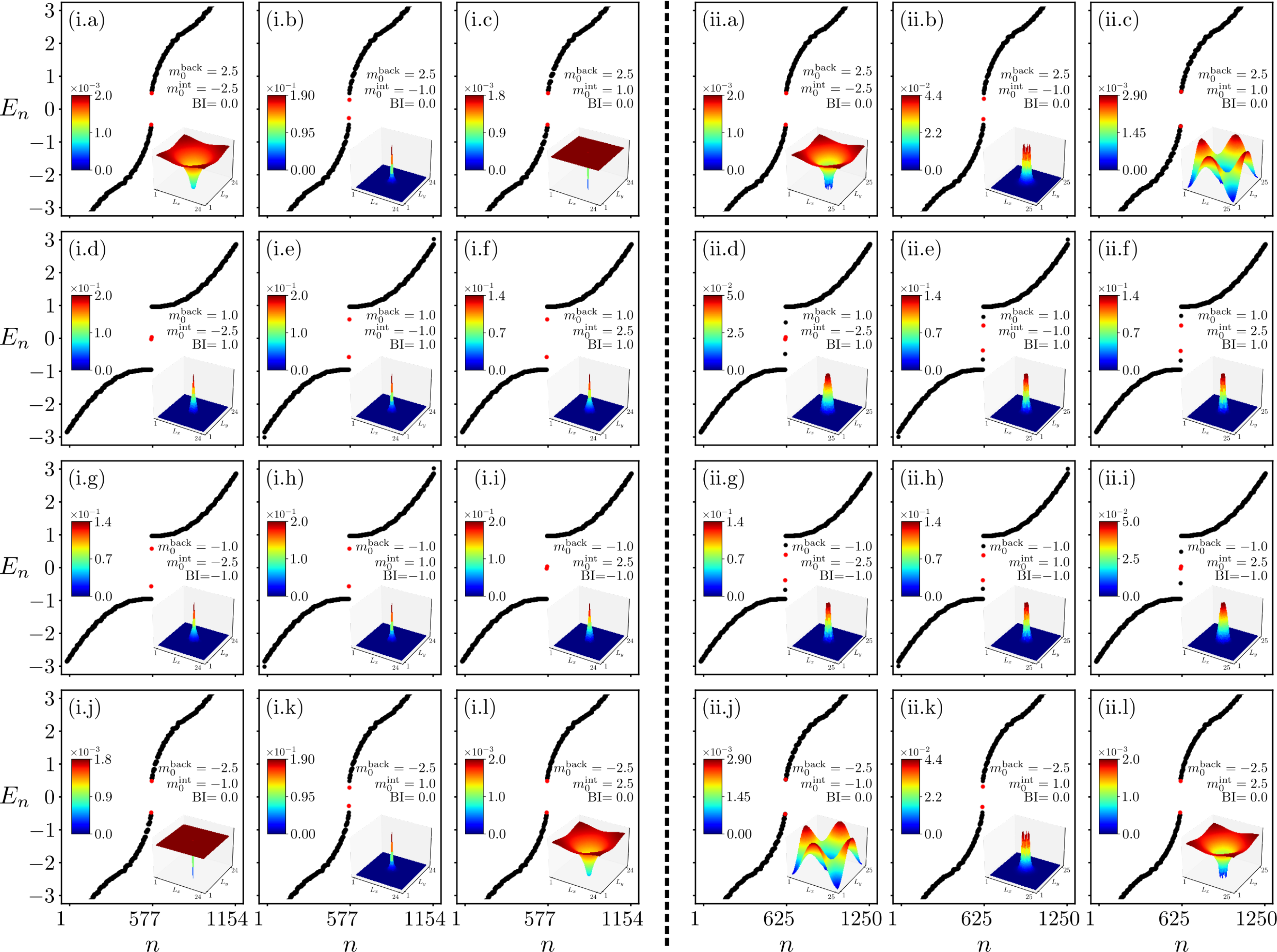}
    \caption{
             Energy eigenvalues ($E_n$) as a function of its index ($n$) in the presence of a single interstitial [(i.a)-(i.l)] and a single Frenkel pair [(ii.a)-(ii.l)] located at the center of a square lattice with linear dimensions $L=24 a$ and $L=25 a$, respectively, in both directions, where $a$ is the lattice spacing. In both cases, we impose periodic boundary conditions in both directions. Inset in each panel shows the joint local density of states for two particle-hole symmetric closest-to-zero-energy modes (shown in red). The corresponding value of the background mass parameter $m^{\rm back}_0$ and the mass parameter at the interstitial site $m^{\rm int}_0$ in the Qi-Wu-Zhang model, respectively represented by the solid and dashed circles (red and blue) in Figs.~\ref{fig:1}(d) and~\ref{fig:1}(e), and the associated Bott index (BI) are quoted in the legend of each panel. See Sec.~\ref{subsec:interstitialtheory} and Sec.~\ref{subsec:frenkeltheory} for a detailed discussion on the results related to the interstitial and Frenkel pair defects, respectively.       
    }
    \label{fig:5}
\end{figure*}
%%%%%%%%%%%%%%%%%%%%%%%%%%%%%%%%%%%%%%%%%%%%%%%%%%%%%%%%%%%%%%%%%%%%%%
%%%%%%%%%%%%%%%%%%%%%%%%%%%%%%%%%%%%%%%%%%%%%%%%%%%%%%%%%%%%%%%%%%%%%%
%%%%%%%%%%%%%%%%%%%%%%%%%%%%%%%%%%%%%%%%%%%%%%%%%%%%%%%%%%%%%%%%%%%%%%

              \subsection{Vacancies}~\label{subsec:vacancytheory}

A vacancy on a square lattice-based model is created by removing both the orbitals from a given site, see Fig.~\ref{fig:1}(a). We choose the linear dimension of the square lattice $L$ in each direction to be an \emph{odd} multiple of $a$, so that the center site(s) of the lattice can subsequently be removed to create a vacancy therein. Although the outcomes, obtained by eliminating a site from the square lattice-based tight-binding model from Eq.~\eqref{eq:realspaceTB}, are insensitive to the choice of the vacancy site. The results are shown in Fig.~\ref{fig:2} and discussed next.

%%%%%%%%%%%%%%%%%%%%%%%%%%%%%%%%%%%%%%%%%%%%%%%%%%%%%%%%%%%%%%%%%%%%%%
%%%%%%%%%%%%%%%%%%%%%%%%%%%%%%%%%%%%%%%%%%%%%%%%%%%%%%%%%%%%%%%%%%%%%%
%%%%%%%%%%%%%%%%%%%%%%%%%%%%%%%%%%%%%%%%%%%%%%%%%%%%%%%%%%%%%%%%%%%%%%
\begin{figure}[t!]
    \centering
    \includegraphics[width=1.00\linewidth]{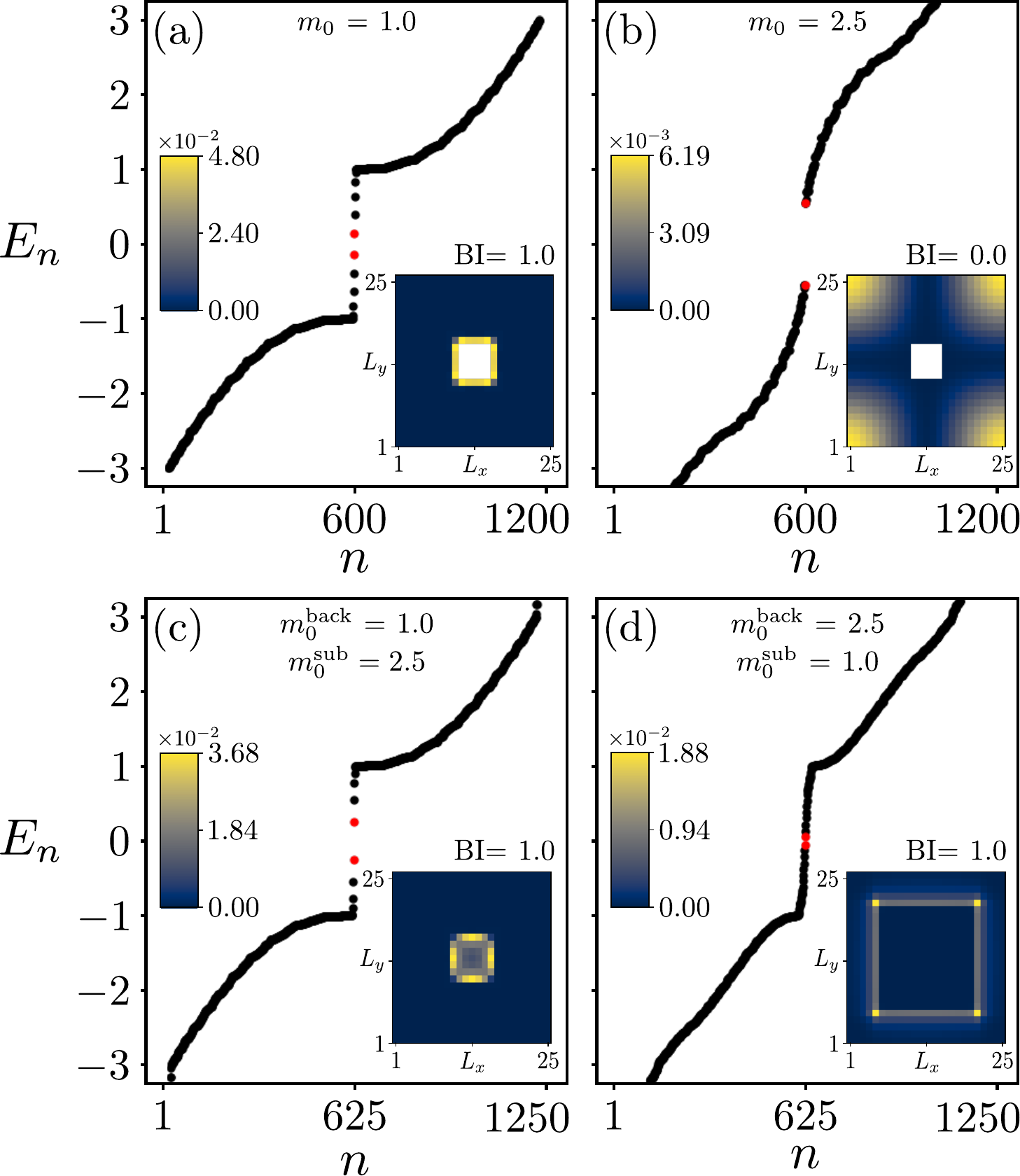}
    \caption{
             Energy eigenvalues ($E_n$) as a function of its index ($n$) with a big vacancy of linear dimension $5a$ in each direction embedded within an original square lattice of linear dimension $25a$ in each direction for (a) $m_0=1.0$ and (b) $m_0=2.5$ with the corresponding inset showing the local density of states for two closest to zero energy modes in each case. Thus, only when the mass parameter $m_0$ is in the topological regime [as in (a)], we find inner edge modes, which are protected by nontrivial Bott index (BI).
             Energy eigenvalues ($E_n$) as a function of its index ($n$) with a big substitution of linear dimension (c) $5a$ in each direction, $m^{\rm back}_0=1.0$, and $m^{\rm sub}_0=2.5$ and (d) $17a$ in each direction, $m^{\rm back}_0=2.5$, and $m^{\rm sub}_0=1.0$ embedded in a parent square lattice of linear dimension $25 a$ in each direction. In both cases, we find the existence of inner edge modes localized around the boundaries of the big substitution, as shown from the local density of states for two closest two zero energy modes in the corresponding inset that are protected by nontrivial Bott index (BI).   
    }
    \label{fig:explanationtheory}
\end{figure}
%%%%%%%%%%%%%%%%%%%%%%%%%%%%%%%%%%%%%%%%%%%%%%%%%%%%%%%%%%%%%%%%%%%%%%
%%%%%%%%%%%%%%%%%%%%%%%%%%%%%%%%%%%%%%%%%%%%%%%%%%%%%%%%%%%%%%%%%%%%%%
%%%%%%%%%%%%%%%%%%%%%%%%%%%%%%%%%%%%%%%%%%%%%%%%%%%%%%%%%%%%%%%%%%%%%%

We consider two configurations for vacancies, one with a single vacancy defect and another one with five such defects. Irrespective of configurations, there exists no in-gap bound state near such defects with a prominent peak in the corresponding local density of states (LDOS) or probability density for $|m_0|>2$ yielding a trivial insulator. On the other hand, when $|m_0|<2$, we find a pair of mid-gap states that are clearly separated from the bulk states with their LDOS peak near the defect even with a single vacancy. With five vacancy sites we observe the emergence of a few more mid-gap bound states near the defect. These observations are qualitatively similar in both $\Gamma$ and ${\rm M}$ phases, which can be justified in the following way. With an increasing number of vacancy sites, the system gradually develops edges inside the system which ultimately host one-dimensional gapless chiral edge modes~\cite{vacancy:1, vacancy:2, vacancy:3, vacancy:4}. Nonetheless, we find that even a single or a few vacancies are sufficient to distinguish between topological and trivial insulators in terms of the vacancy-localized mid-gap states. We also confirm that the value of the Bott index does not change, at least with the introduction of such a small numbers of vacancies in the interior of the system. We note that the vacancy-bound topological in-gap modes with their LDOS peak around such defects are immune to a small amount of random charge impurities, see Appendix~\ref{appensec:disorder}. Finally, we notice that such bound states always maintain a finite gap with the rest of the bulk states except at the points of topological band gap closing at $m_0/t_0=0,\pm 2$ (not shown explicitly).

%%%%%%%%%%%%%%%%%%%%%%%%%%%%%%%%%%%%%%%%%%%%%%%%%%%%%%%%%%%%%%%%%%%%%%
%%%%%%%%%%%%%%%%%%%%%%%%%%%%%%%%%%%%%%%%%%%%%%%%%%%%%%%%%%%%%%%%%%%%%%
%%%%%%%%%%%%%%%%%%%%%%%%%%%%%%%%%%%%%%%%%%%%%%%%%%%%%%%%%%%%%%%%%%%%%%
\begin{figure*}[t!]
    \centering
    \includegraphics[width=1.00\linewidth]{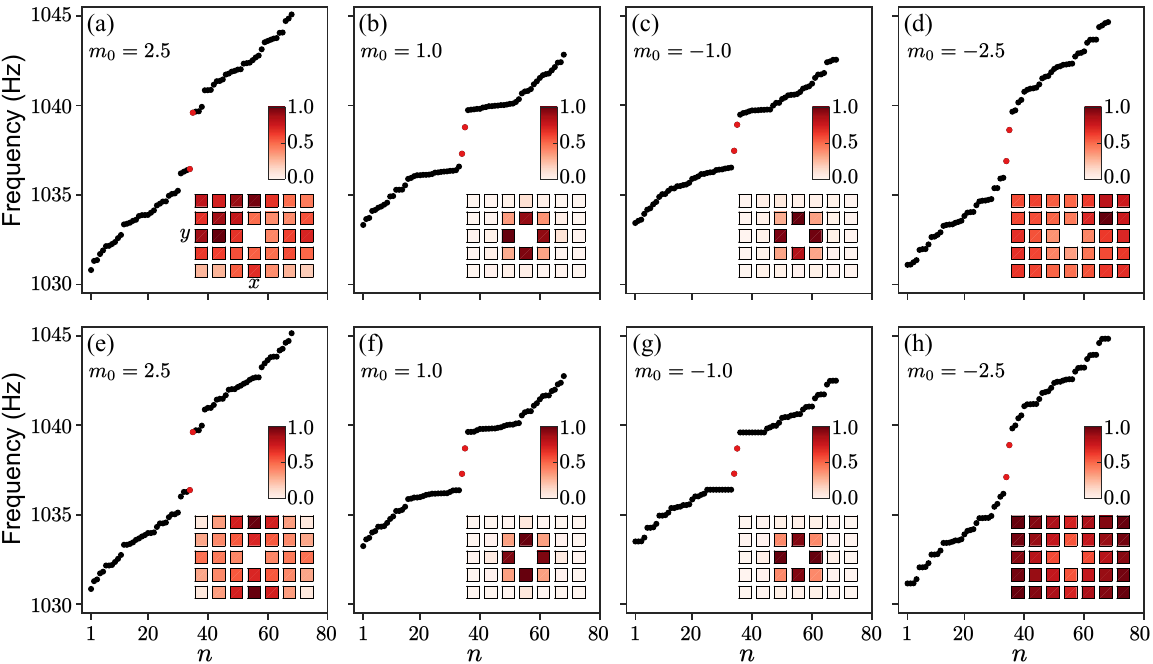}
    \caption{
        Experimental observations [(a)-(d)] and the corresponding simulations [(e)-(h)] of the energy eigenvalues (frequencies) as a function of its index ($n$) in the presence of a single vacancy located at the center of a $7\times 5$ lattice with periodic boundary conditions in both directions. Inset in each panel shows the joint local density of states (normalized by its maximal value) for two particle-hole symmetric closest-to-zero-energy modes or equivalently for the modes symmetrically placed about and closest to 1038 Hz (shown in red). The mass parameter value ($m_0$) is quoted in each panel. Compare with Fig.~\ref{fig:2} and see Sec.~\ref{subsec:vacancyexp} for details. 
    }
    \label{fig:exp_vacancy}
\end{figure*}
%%%%%%%%%%%%%%%%%%%%%%%%%%%%%%%%%%%%%%%%%%%%%%%%%%%%%%%%%%%%%%%%%%%%%%
%%%%%%%%%%%%%%%%%%%%%%%%%%%%%%%%%%%%%%%%%%%%%%%%%%%%%%%%%%%%%%%%%%%%%%
%%%%%%%%%%%%%%%%%%%%%%%%%%%%%%%%%%%%%%%%%%%%%%%%%%%%%%%%%%%%%%%%%%%%%%

              \subsection{Schottky defect}~\label{subsec:schottkytheory}

The Schottky defect is a special type of lattice defect that can only be realized in diatomic crystals, such as NaCl. Such a defect is realized by removing sodium (Na) and chlorine (Cl) ions with positive and negative charges, respectively, at different sites (otherwise at arbitrary locations) that locally destroys the charge neutrality while maintaining it globally. In the context of the QWZ model, a Schottky pair defect can be realized in the following way. Ions with the positive and negative charges in NaCl are now replaced by two orbitals with opposite parity eigenvalues $+1$ and $-1$. With such a one-to-one correspondence in hand, the rest of the construction for the Schottky defect follows straightforwardly, as shown in Fig.~\ref{fig:1}(b). The corresponding tight-binding model is constructed from $H^{\rm SL}_{\rm TB}$ [Eq.~\eqref{eq:realspaceTB}] upon eliminating two orbitals from two distinct sites. Therefore, a Schottky defect locally develops a boundary within the systems but at two sites for opposite orbitals, which we name orbital-selective or half boundary.

We find that the outcomes in terms of the presence or absence of any bound state around such a defect in the topological or trivial parameter regime are qualitatively similar to the ones we previously found with a single vacancy defect, as shown in Fig.~\ref{fig:3} (top row). To underpin the role of vacancies for complementary orbitals at different sites of the lattice, yielding the Schottky defect, we perform the same numerical analysis by spatially separating two such partial vacancy sites. The results are shown in Fig.~\ref{fig:3} (bottom row). When the two sites of the Schottky defect are spatially separated, we find prominent LDOS peaks on both the sites, implying that orbital-selective or half-boundary can individually host topological bound modes irrespective of their locations. Therefore, Schottky defects are also instrumental in identifying topological insulators from their trivial counterparts. Finally, we note that the value of the Bott index does not change from the one in a perfect square lattice model, at least when the number of Schottky defects is sufficiently small.  Also, the mid-gap modes bound to the Schottky defect are immune to a small amount of random charge impurities (see Appendix~\ref{appensec:disorder}).

              \subsection{Substitutions}~\label{subsec:substitutiontheory}

Next we focus on a different type of non-geometric lattice defects, substitutions, that does not involve removal of any site from the system. Traditionally, a substitution defect is created by changing the local chemical environment at a regular atomic site in a crystal. In the context of the QWZ model, such defects are created by substituting the on-site staggered mass $m_0$ at a given set of sites to a different value $m^{\rm sub}_0$ (say) in comparison to that in the rest of the sites, constituting its background value $m^{\rm back}_0$ (say), as shown in Fig.~\ref{fig:1}(c). We always choose the pair of staggered mass parameters $(m^{\rm back}_0,m^{\rm sub}_0)$ in two distinct parameter regimes, yielding altogether \emph{twelve} configurations (since the QWZ model yields four distinct parameter regimes, see Sec.~\ref{sec:model}). The corresponding tight-binding Hamiltonian is constructed from $H^{\rm SL}_{\rm TB}$ with $m_0=m^{\rm back}_0$ by taking $m_0 \to m^{\rm sub}_0$ at the site(s) of the substitution defect. Additionally, we study single-site and five-site substitution defects that are created near the center of the square lattice. For each of such mass parameter set, we numerically search for topologically robust mid-gap modes near the substitution defects. The results are shown in Fig.~\ref{fig:4}, which we discuss next.

%%%%%%%%%%%%%%%%%%%%%%%%%%%%%%%%%%%%%%%%%%%%%%%%%%%%%%%%%%%%%%%%%%%%%%
%%%%%%%%%%%%%%%%%%%%%%%%%%%%%%%%%%%%%%%%%%%%%%%%%%%%%%%%%%%%%%%%%%%%%%
%%%%%%%%%%%%%%%%%%%%%%%%%%%%%%%%%%%%%%%%%%%%%%%%%%%%%%%%%%%%%%%%%%%%%%
\begin{figure*}[t!]
    \centering
    \includegraphics[width=1.00\linewidth]{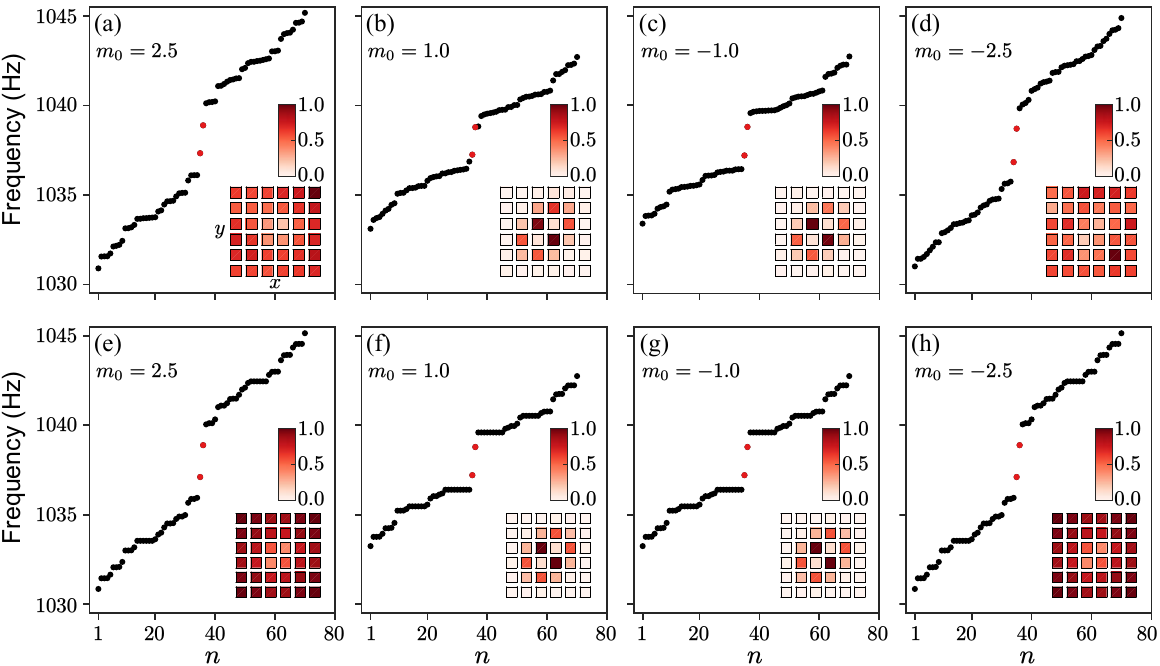}
    \caption{
         Experimental observations [(a)-(d)] and the corresponding simulations [(e)-(h)] of the energy eigenvalues (frequencies) as a function of its index ($n$) in the presence of a single Schottky defect, where the vacancies of opposite-parity orbitals are separated by a distance of $\sqrt{2}a$, where $a$ is the lattice spacing in the $x$ and $y$ directions. The system consists of a $6\times6$ lattice with periodic boundary conditions applied along both directions. Inset in each panel shows the joint local density of states (normalized by its maximal value) for two particle-hole symmetric closest-to-zero-energy modes or equivalently for the modes symmetrically placed about and closest to 1038 Hz (shown in red). The mass parameter value ($m_0$) is quoted in each panel. Compare with Fig.~\ref{fig:3} and see Sec.~\ref{subsec:schottkyexp} for details.  
    }
    \label{fig:exp_Schottky}
\end{figure*}
%%%%%%%%%%%%%%%%%%%%%%%%%%%%%%%%%%%%%%%%%%%%%%%%%%%%%%%%%%%%%%%%%%%%%%
%%%%%%%%%%%%%%%%%%%%%%%%%%%%%%%%%%%%%%%%%%%%%%%%%%%%%%%%%%%%%%%%%%%%%%
%%%%%%%%%%%%%%%%%%%%%%%%%%%%%%%%%%%%%%%%%%%%%%%%%%%%%%%%%%%%%%%%%%%%%%

First, we lock $m^{\rm back}_0$ within topologically trivial parameter regimes ($m_0>2$ and $m_0<-2$). If then $m^{\rm sub}_0$ as well falls within the trivial parameter regime there exists no mid-gap bound states with a LDOS peak around the substitution site(s). By contrast, with $m^{\rm sub}_0$ falling within the topological parameter regimes, bound states with their highly peaked LDOS around the defect site(s) appear in the spectrum. With increasing number of substitution sites in a given region of the system, such modes gradually get decoupled from the rest of the bulk states, which can be seen by comparing the results from Fig.~\ref{fig:4} with single-site (left panel) and five-site (right panel) substitution. This outcome can be justified in the following way. With a background trivial insulator, even when a small number of sites are placed in the topological regime, a boundary between a local topological and global trivial insulators is created, hosting such robust mig-gap states, bound to the defect.

On the other hand, when we lock $m^{\rm back}_0$ to one of the topological parameter regimes and choose $m^{\rm sub}_0$ from any one of three other parameter regime, there always exists mid-gap bound states well-separated from the bulk states with their LDOS being highly peaked around the substitution sites. This outcome can be appreciated in the following way. With the above mentioned substitution protocol with a fixed topological $m^{\rm back}_0$, the system develops a local boundary with either a trivial or distinct topological insulator, thereby binding robust mid-gap modes in its vicinity. The number of such mid-gap modes increases with increasing number of substitution sites, as the system then gradually develops one-dimensional chiral edge modes within the system. Even in this case, we find that the value of the Bott index matches with the one for $m_0=m^{\rm back}_0$ in the QWZ model, at least when the number of substitution sites is sufficiently small in comparison to the total number of sites of the parent square lattice. The mid-gap modes bound to the substitution defect are also stable against random point-like charge impurities (see Appendix~\ref{appensec:disorder}).

%%%%%%%%%%%%%%%%%%%%%%%%%%%%%%%%%%%%%%%%%%%%%%%%%%%%%%%%%%%%%%%%%%%%%%
%%%%%%%%%%%%%%%%%%%%%%%%%%%%%%%%%%%%%%%%%%%%%%%%%%%%%%%%%%%%%%%%%%%%%%
%%%%%%%%%%%%%%%%%%%%%%%%%%%%%%%%%%%%%%%%%%%%%%%%%%%%%%%%%%%%%%%%%%%%%%
\begin{figure*}[t!]
    \centering
    \includegraphics[width=1.00\linewidth]{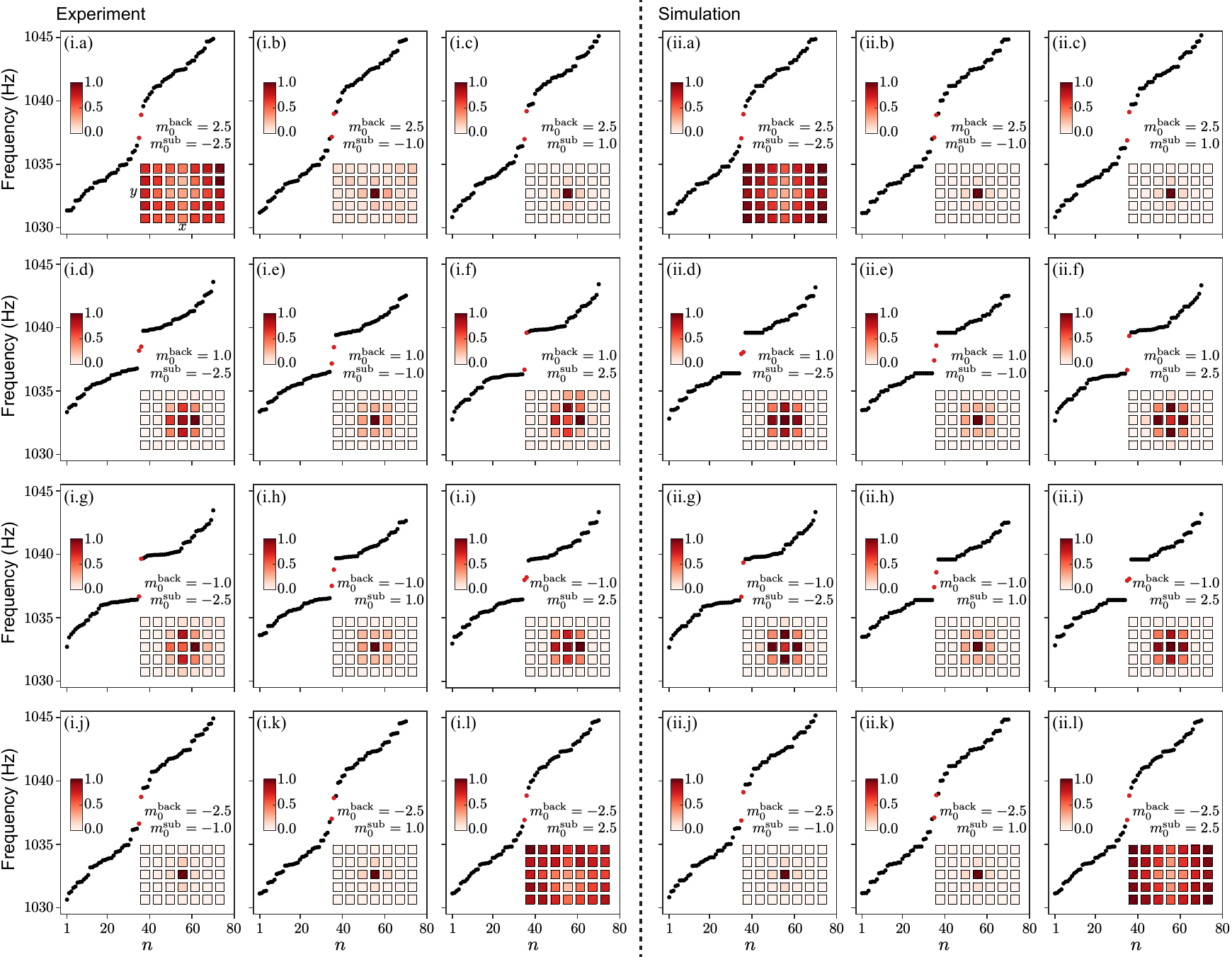}
    \caption{
        (i) Experimental observations and (ii) the corresponding simulations of the energy eigenvalues (frequencies) as a function of its index ($n$) in the presence of a single substitution located at the center of a $7\times 5$ lattice with periodic boundary conditions in both directions. Inset in each panel shows the joint local density of states (normalized by its maximal value) for two particle-hole symmetric closest-to-zero-energy modes or equivalently for the modes symmetrically placed about and closest to 1038 Hz (shown in red). The mass parameter values $m^{\rm back}_0$ and $m^{\rm sub}_0$ are quoted in each panel. Compare with Fig.~\ref{fig:4} and see Sec.~\ref{subsec:substitutionexp} for details. 
    }
    \label{fig:exp_substitution}
\end{figure*}
%%%%%%%%%%%%%%%%%%%%%%%%%%%%%%%%%%%%%%%%%%%%%%%%%%%%%%%%%%%%%%%%%%%%%%
%%%%%%%%%%%%%%%%%%%%%%%%%%%%%%%%%%%%%%%%%%%%%%%%%%%%%%%%%%%%%%%%%%%%%%
%%%%%%%%%%%%%%%%%%%%%%%%%%%%%%%%%%%%%%%%%%%%%%%%%%%%%%%%%%%%%%%%%%%%%%

              \subsection{Interstitial}~\label{subsec:interstitialtheory}

Traditionally, an interstitial defect occurs when an \emph{extra} atom occupies a position which does not belong to the set of regular atomic sites of the underlying periodic crystal, see Fig.~\ref{fig:1}(d). Next, we show that such a lattice defect can also be instrumental in probing both changes in the local topological environment in a normal insulator and topological crystals. For this purpose, we place a site at an irregular point, where the value of $m_0$, also denoted by $m^{\rm int}_0$, is chosen from a topologically distinct parameter regime of the QWZ model, in comparison to that on the parent square lattice, which is denoted by $m^{\rm back}_0$. Thus, with an interstitial we consider twelve possible parameter values, determined by $(m^{\rm back}_0, m^{\rm int}_0)$. As mentioned previously, in order to ensure that the atomic orbitals at such an irregular site are properly connected to its surrounding regular sites, we now employ the generalized tight-binding model from Eq.~\eqref{eq:Method1} in the presence of an interstitial. The results are shown in Fig.~\ref{fig:5} (left column).

As an interstitial can qualitatively be considered as a substitution, but at an irregular lattice site, the results are qualitatively similar to the ones we discussed previously with substitution defects, at least when $m^{\rm back}_0$ belongs to one of the topological parameter regimes. By contrast, when $m^{\rm back}_0$ and $m^{\rm int}_0$ belong to one of the trivial and topological parameter regimes, respectively, such that ${\bf K}_{\rm min}$ with $m_0=m^{\rm back}_0$ and ${\bf K}_{\rm inv}$ with $m_0=m^{\rm int}_0$ in the QWZ model are at the same points in the FBZ, as is the case with $m^{\rm back}_0=2.5$ and $m^{\rm int}_0=1.0$ (see Sec.~\ref{sec:model}), for example, there exists no mid-gap mode that is highly localized near the interstitial defect. This outcome is in stark contrast to the situation with a substitution defect for the same set of parameters (compare with Fig.~\ref{fig:4}) and can be supported in the following way. Notice that introduction of an interstitial site leads to short-distance or large-momentum scattering, as it is placed in-between the regular lattice sites. When $m^{\rm back}_0$ is in the trivial regime with a given ${\bf K}_{\rm min}$ and $m^{\rm int}_0$ is in the topological regime with a given ${\bf K}_{\rm inv}$, such that ${\bf K}_{\rm min}={\bf K}_{\rm inv}$ the local short-distance or large-momentum scattering in the background of a trivial insulators due to the presence of a single topological interstitial site is insufficient to change the local topological environment to bind any robust mode therein. Such a short-distance or large-momentum scattering, nonetheless, binds robust modes when the difference between ${\bf K}_{\rm min}$ and ${\bf K}_{\rm inv}$ is also large. Finally, we note that even in the presence of an interstitial the Bott index remains the same as in a pristine square lattice model with $m_0 \equiv m^{\rm back}_0$. Otherwise, all the mid-gap modes bound to interstitial are robust against weak random on-site charge impurities as discussed in Appendix~\ref{appensec:disorder}.

%%%%%%%%%%%%%%%%%%%%%%%%%%%%%%%%%%%%%%%%%%%%%%%%%%%%%%%%%%%%%%%%%%%%%%
%%%%%%%%%%%%%%%%%%%%%%%%%%%%%%%%%%%%%%%%%%%%%%%%%%%%%%%%%%%%%%%%%%%%%%
%%%%%%%%%%%%%%%%%%%%%%%%%%%%%%%%%%%%%%%%%%%%%%%%%%%%%%%%%%%%%%%%%%%%%%
\begin{figure*}[t!]
    \centering
    \includegraphics[width=1.00\linewidth]{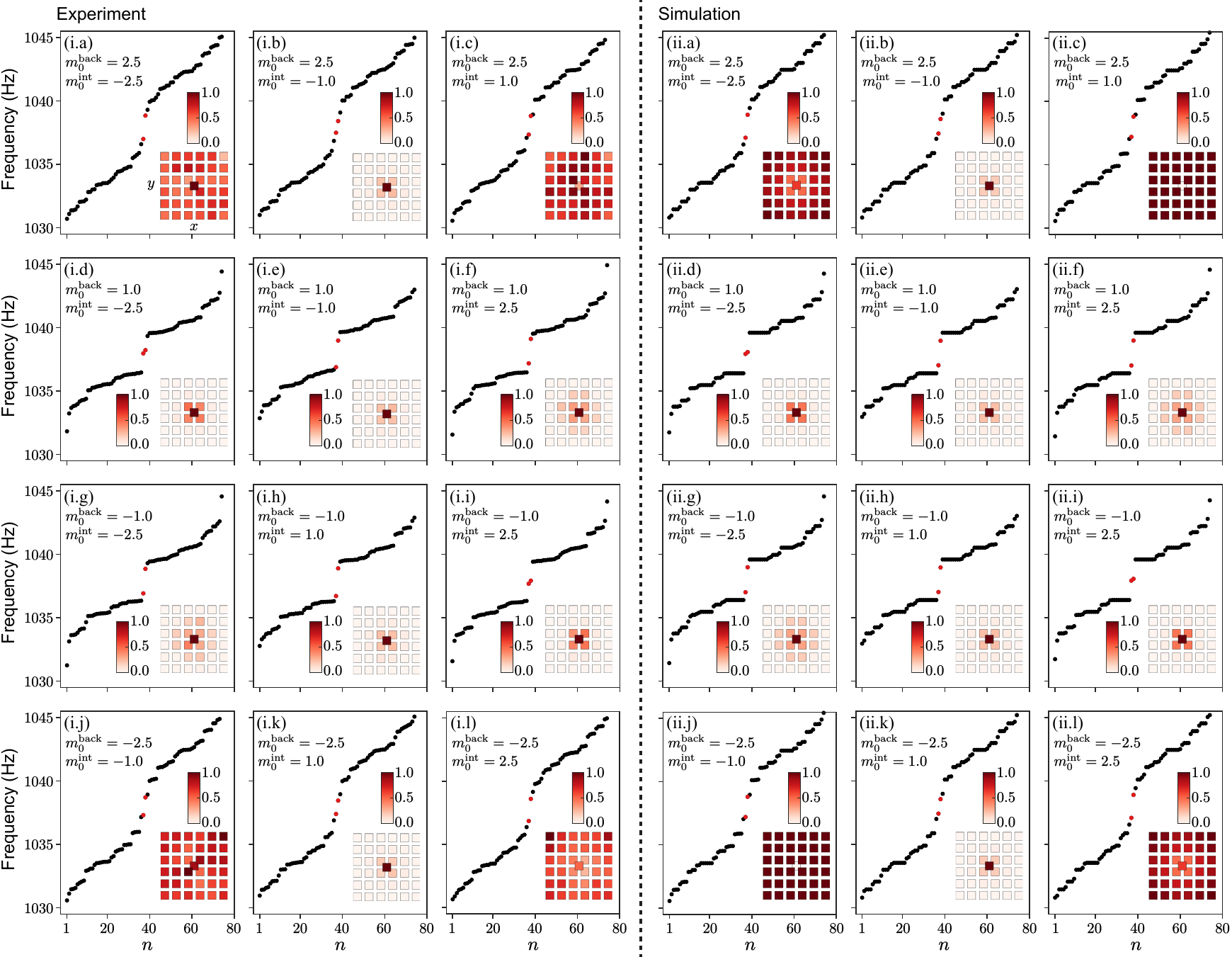}
    \caption{
        (i) Experimental observations and (ii) the corresponding simulations of the energy eigenvalues (frequencies) as a function of its index ($n$) in the presence of a single interstitial located at the center of a $6\times 6$ lattice with periodic boundary conditions in both directions. Inset in each panel shows the joint local density of states (normalized by its maximal value) for two particle-hole symmetric closest-to-zero-energy modes or equivalently for the modes symmetrically placed about and closest to 1038 Hz (shown in red). The mass parameter values $m^{\rm back}_0$ and $m^{\rm int}_0$ are quoted in each panel. Compare with Fig.~\ref{fig:5} (left column) and see Sec.~\ref{subsec:interstitialexp} for details. 
    }
    \label{fig:exp_interstitial}
\end{figure*}
%%%%%%%%%%%%%%%%%%%%%%%%%%%%%%%%%%%%%%%%%%%%%%%%%%%%%%%%%%%%%%%%%%%%%%
%%%%%%%%%%%%%%%%%%%%%%%%%%%%%%%%%%%%%%%%%%%%%%%%%%%%%%%%%%%%%%%%%%%%%%
%%%%%%%%%%%%%%%%%%%%%%%%%%%%%%%%%%%%%%%%%%%%%%%%%%%%%%%%%%%%%%%%%%%%%%

              \subsection{Frenkel pair}~\label{subsec:frenkeltheory}

Finally, we consider a Frenkel pair defect, shown in Fig.~\ref{fig:1}(e). A Frenkel pair can be considered as a superposition of a vacancy [Fig.~\ref{fig:1}(a)] and an interstitial [Fig.~\ref{fig:1}(d)]. The corresponding tight-binding Hamiltonian is constructed from $H^{\rm SL, gen}_{\rm TB}$ [Eq.~\eqref{eq:Method1}] after removing a site from a regular atomic sites of the square lattice (yielding a vacancy) and introducing one site at an irregular point (producing an interstitial). Therefore, we parameterize a Frenkel pair by a pair of mass parameter values $m^{\rm back}_0$ and $m^{\rm int}_0$, respectively, denoting the value of $m_0$ in the QWZ model on the sites of the parent square lattice and at the interstitial site. Therefore, we once again scrutinize twelve configurations, and the corresponding results are shown in Fig.~\ref{fig:5} (right column).

To this end, we find that the outcomes with a Frenkel pair defect in terms of the presence or absence of defect-bound mid-gap states are qualitatively similar to the ones we previously reported with an interstitial. This observation can be justified in the following way. Vacancy-bound modes exists only when $m^{\rm back}_0$ belongs to the topological parameter regime (see Fig.~\ref{fig:2}), which remains stable with an interstitial defect for any arbitrary choice of $m^{\rm int}_0$, as shown in Fig.~\ref{fig:5} (left column). On the other hand, an interstitial can bind in-gap defect modes even when $m^{\rm back}_0$ is within the trivial parameter regime, depending on the value of $m^{\rm int}_0$, discussed in the last subsection. Consequently, for a Frenkel pair (a superposition of a vacancy and an interstitial), we find similar outcomes as with a single interstitial. Namely, when mid-gap modes exist with $m^{\rm back}_0$ falling within the trivial parameter regime, its LDOS is somewhat exclusively localized on the interstitial site. On the other hand, when mid-gap modes exist with $m^{\rm back}_0$ falling within the topological parameter regime, its LDOS is distributed between the interstitial site and sites near the vacancy. We note that mid-gap modes bound to a Frenkel pair are stable against sufficiently weak on-site random charge impurities, which we discuss in Appendix~\ref{appensec:disorder}. And the Bott index does not change with sufficiently small number of such defect.

%%%%%%%%%%%%%%%%%%%%%%%%%%%%%%%%%%%%%%%%%%%%%%%%%%%%%%%%%%%%%%%%%%%%%%
%%%%%%%%%%%%%%%%%%%%%%%%%%%%%%%%%%%%%%%%%%%%%%%%%%%%%%%%%%%%%%%%%%%%%%
%%%%%%%%%%%%%%%%%%%%%%%%%%%%%%%%%%%%%%%%%%%%%%%%%%%%%%%%%%%%%%%%%%%%%%
\begin{figure*}[t!]
    \centering
    \includegraphics[width=1.00\linewidth]{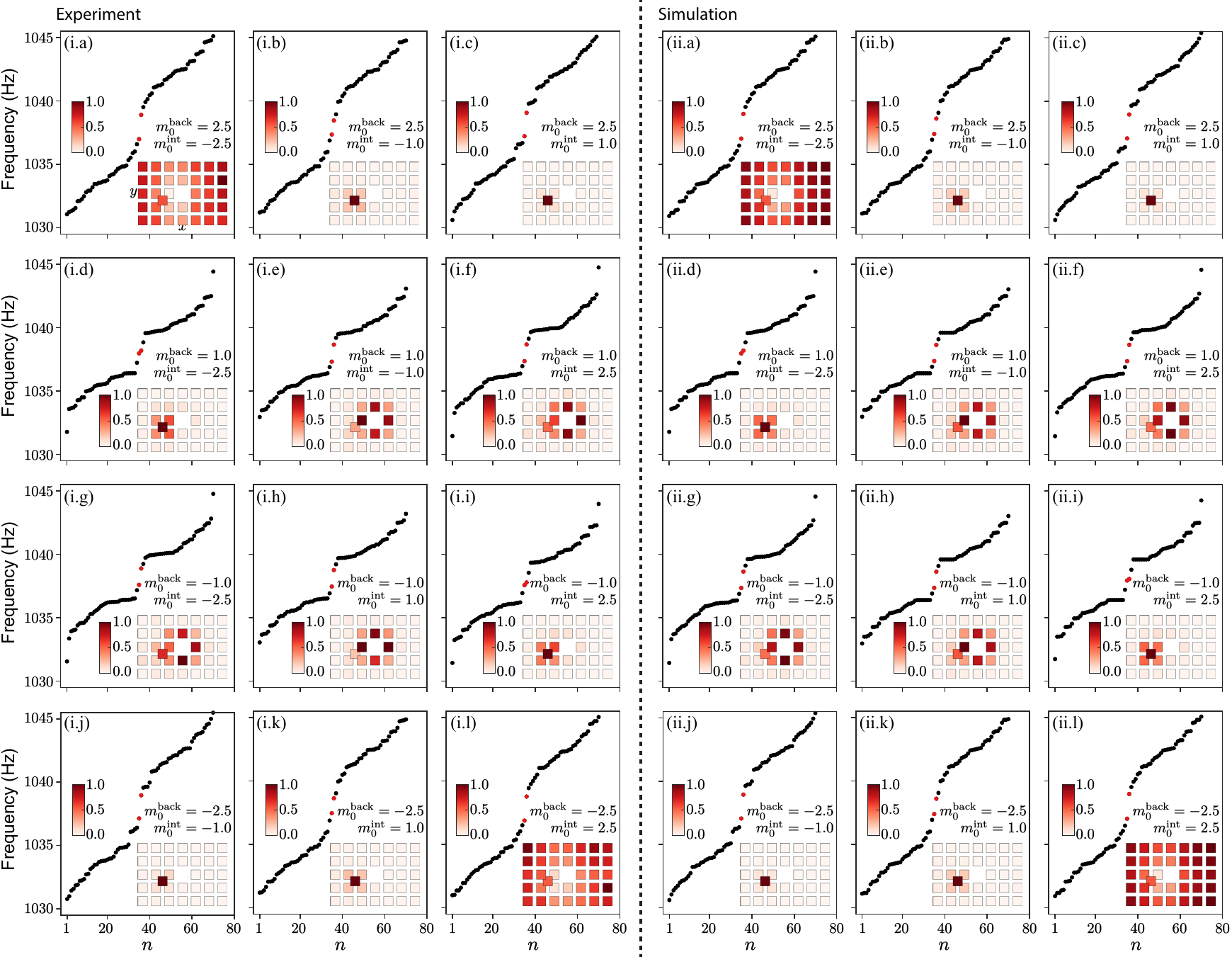}
    \caption{
        (i) Experimental observations and (ii) the corresponding simulations of the energy eigenvalues (frequencies) as a function of its index ($n$) in the presence of a single Frenkel pair located near the center of a $7\times 5$ lattice with periodic boundary conditions in both directions. Inset in each panel shows the joint local density of states for two particle-hole symmetric closest-to-zero-energy modes or equivalently for the modes symmetrically placed about and closest to 1038 Hz (shown in red). The mass parameter values $m^{\rm back}_0$ and $m^{\rm int}_0$ are quoted in each subfigure. Compare with Fig.~\ref{fig:5} (right column) and see Sec.~\ref{subsec:frenkelpairexp} for details.  
    }
    \label{fig:exp_Frenkel}
\end{figure*}
%%%%%%%%%%%%%%%%%%%%%%%%%%%%%%%%%%%%%%%%%%%%%%%%%%%%%%%%%%%%%%%%%%%%%%
%%%%%%%%%%%%%%%%%%%%%%%%%%%%%%%%%%%%%%%%%%%%%%%%%%%%%%%%%%%%%%%%%%%%%%
%%%%%%%%%%%%%%%%%%%%%%%%%%%%%%%%%%%%%%%%%%%%%%%%%%%%%%%%%%%%%%%%%%%%%%

\subsection{Topological origin of defect modes}

With the existence of defect-bound localized modes being established theoretically, we now propose a possible topological origin of them. In this context, we note the following. (a) Consider a topological insulator with a big vacancy near the center of the system with periodic boundary conditions applied on its outer edges. Such a system supports edge modes near the inner edges, which is protected by a bulk invariant (Bott index), as shown in Fig.~\ref{fig:explanationtheory}(a). Now as the size of vacancy gets smaller the edge modes gradually start to hybridize and eventually with a single vacancy or a small number of vacancies, there exists only a few in-gap modes. For a same geometry, however, with an underlying normal or trivial insulator, there exists no in-gap mode irrespective of the dimensionality of the vacancy, when the bulk invariant is also trivial, as shown in Fig.~\ref{fig:explanationtheory}(b). (b) Now with a topological (normal) insulator fostering a chunk of substituted normal (topological) insulator near its center, the edge modes appear near their junction that are also protected by a nontrivial bulk topological invariant, as shown in Figs.~\ref{fig:explanationtheory}(c) and~\ref{fig:explanationtheory}(d). With a gradual shrinking of such a substitution, the edge modes begin to hybridize, and finally with a single substitution or a small number of substitutions, there exists a pair of or a small number of in-gap modes. With these observations in hand, we conclude that in-gap bound states near (i) vacancy and Schottky defects are supported by (a), and (ii) substitution, interstitial, and Frenkel pair are supported by (b). It should, however, be noted that with pointlike substitution, interstitial, and Frenkel pair defects with $m^{\rm back}_0$ ($m^{\rm sub}_0$) in the trivial (topological) regime, the Bott index vanishes, but the mid-gap modes continue to survive as they originally stemmed from topological edge modes. Therefore, all the modes we discussed and found in this work theoretically so far and experimentally, which we discuss next, are of topological origin.

\section{Experimental setups on acoustic lattices: Measurements and validations}~\label{sec:experiment}

In this section, we set out to experimentally validate the prominent theoretical predictions, discussed in the last section. For this purpose, we implement the QWZ operator, namely ${\boldsymbol \tau} \cdot \vec{d}(\vec{k})$, in an active acoustic lattice via Fourier transformation, where the acoustic cavities are used to emulate orbitals and lattice sites, while microphones and speakers are used to realize hopping amplitudes. The tight-binding analogy of such networks makes them ideal for engineering staggered onsite potentials and hopping amplitudes on a highly tunable classical meta-material platform. See Fig.~\ref{fig:exp_setup}.

In our design, each acoustic cavity supports a dipole resonant mode measured at a complex frequency $\omega_0 = 1038\,\mathrm{Hz}-6i \; \mathrm{Hz}$, where the imaginary component accounts for the intrinsic background loss. In experiments, all the hopping parameters and the staggered potential are identical to those used in the numerical simulations but scaled by a factor of $s$. Consequently, the experimentally realized Hamiltonian is given by $H^\mathrm{exp} = \omega_0 + s H^\mathrm{sim}$, where $H^\mathrm{sim}$ represents the corresponding Hamiltonian used in the simulations.
The scale factor $s$ cannot be too small, otherwise the eigenvalues of $H^{\mathrm{exp}}$ will be too close to $\omega_0$, making it difficult to resolve spectral features and accurately extract the eigenvalues and eigenvectors from the measured Green's functions. On the other hand, if $s$ is too large, the eigenfrequencies shift too far from $\omega_0$, the resonant frequency of the acoustic cavities, which leads to reduced signal strength and increased noise due to the finite resonator bandwidth. In experiments, we find that $s=1.6$ provides an optimal balance between these competing factors and yields improved measurement accuracy.

For a given real-space QWZ lattice containing any type of defect, the system is described by a Hamiltonian $H^\mathrm{exp}$. The diagonal elements of $H^\mathrm{exp}$ represent the staggered on-site potentials, corresponding to the shifted resonant frequencies of the acoustic cavities. Such staggered potentials are implemented through a self-feedback loop formed by a microphone and a speaker installed within the same cavity~\cite{dislocation:19}. Each nonzero off-diagonal element, such as $H^\mathrm{exp}_{j,k}$, represents a unidirectional hopping from site $k$ to site $j$. As illustrated in Fig.~\ref{fig:exp_setup}(a), these hopping terms are realized using a microphone–speaker pair: the microphone in cavity $k$ (1) records the local acoustic pressure, which is then amplified and phase-shifted by a controller to drive the speaker in cavity $j$ (2). 
    The effective Hamiltonian reads
\begin{equation}
H =
\begin{pmatrix}
\omega_0 + s m_0 & 0 \\
i s t/2 & \omega_0 - s m_0
\end{pmatrix}.
\end{equation}
Here, the two cavities represent two orbital basis states of the effective tight-binding Hamiltonian. 
We emphasize that this configuration is fundamentally different from the conventional symmetric and antisymmetric supermodes of two reciprocally coupled cavities. 
Instead, each cavity supports one dipole resonant mode, while the two orbital components of the Qi-Wu-Zhang model are encoded through the staggered onsite terms $\omega_0 \pm s m_0$ and the engineered unidirectional hopping.

The effective hopping strength, $ist/2$, is calibrated by fitting the measured transmission response between the two corresponding cavities.
In practice, the effective hopping strength is calibrated through measurement rather than determined solely by the nominal amplifier gain. By measuring the cross-power spectral density between two cavities and extracting the complex ratio $P(\omega)=\psi_2(\omega)/\psi_1(\omega)$, the effective hopping $\kappa_0$ is obtained. The amplifier gain is then iteratively adjusted according to $g^{(n+1)} = (\kappa_0^\mathrm{tgt}/\kappa_0^{(n)}) g^{(n)}$ until the extracted hopping matches the target value within experimental tolerance (approximately 0.1\,Hz), where $n$ denotes the iteration step.
The experimental data and fitted curves in Fig.~\ref{fig:exp_setup}(c) confirm the accuracy and validity of the implemented hopping amplitudes.
All measurements in the present work were performed at room temperature ($23^\circ\mathrm{C}$). The long-term stability of the active acoustic platform has been systematically characterized in Ref.~\cite{dislocation:19}, where the tight-binding parameters were monitored continuously over a six-hour period under sustained operation. This validated stability window exceeds the acquisition time of the present measurements, which is less than three hours for each full Green's-function measurement. Therefore, thermal drift of electrical components does not affect the reported results.

In the experimental setup [Fig.~\ref{fig:exp_setup}(b)], the hopping amplitudes and the phases are controlled by an amplifier and a phase shifter, respectively. While an external phase-shifter module can be used~\cite{Zhang2023ObservationAcousticNonHermitian}, it requires manual adjustment; instead, we integrate this functionality into the controller, enabling flexible and programmable tuning of the hopping phase. A key advantage of this hopping implementation approach is its versatility in breaking time-reversal symmetry, which is essential for realizing the QWZ model. For instance, the hopping term $s t \sin(k_x a)$ can be implemented by adding an additional unidirectional hopping in Fig.~\ref{fig:exp_setup}(a) from Cavity 2 to Cavity 1 with a strength of $-ist/2$.

To construct the lattice, we assemble an array of acoustic cavities serving as lattice sites. Figure~\ref{fig:exp_setup}(d) illustrates a representative $7 \times 5$ lattice with a single vacancy at its center, where the spheres denote the sites and orbitals. Each hopping  (represented by tubes) is individually tuned while all other hoppings are temporarily disabled to avoid interference. After all hoppings are calibrated, they are simultaneously activated for subsequent measurements. An additional advantage of this hopping implementation is that it imposes no spatial constraints on the placement of cavities. As a result, periodic boundary conditions can be directly realized by introducing couplings between boundary sites on the opposite edges.

In conventional measurements, a speaker (pump) is typically used to excite the lattice at a specific site, while microphones (probes) record the acoustic responses at all sites. However, such measurements provide only indirect signatures of bound states and can be contaminated by background losses. In this work, we instead employ a Green's-function-based measurement approach, which enables direct reconstruction of the complete spectra and eigenstates~\cite{Zhong2025ExperimentallyProbingNonHermitian}. The Green's function matrix $G(\omega)$ is directly related to the lattice Hamiltonian via $G(\omega) = (\omega-H^\mathrm{exp})^{-1}$, where $\omega$ is the excitation frequency.
Using the spectral decomposition, $H^\mathrm{exp}= \sum_n E_n \ket{\psi_n} \bra{\psi_n}$, the Green's function can be expressed as $G (\omega) = \sum_n  \ket{\psi_n}\bra{\psi_n}/(\omega-E_n)$. Here, $E_n$ is the eigenvalue, and $\ket{\psi_n}$ is the corresponding eigenstate. When $G(\omega)$ acts on the eigenstate $\ket{\psi_n}$, it yields 
\begin{equation}
G(\omega) \ket{\psi_n}    
=  
\frac{1}{\omega-E_n} 
 \ket{\psi_n}    
 .
\label{eq:green_spec}
\end{equation}
Equation~(\ref{eq:green_spec}) shows that the eigenstates of $G(\omega)$ are identical to those of $H^\mathrm{exp}$, while the eigenvalues of $G(\omega)$ are related to the Hamiltonian eigenvalues $E_n$ through ${1}/({\omega - E_n})$. Consequently, the complete energy spectrum and corresponding eigenstates can be obtained by diagonalizing the experimentally measured Green’s function.

To obtain the full Green's function  $G(\omega)$ in experiments, the pump is sequentially activated at frequencies ranging from 1000\,Hz to 1080\,Hz at each cavity site, and the resulting acoustic pressure is recorded in all cavities using microphones. This procedure is systematically repeated for every site across the entire lattice.
    The uncertainty in the reconstructed energy spectra is negligible under the present experimental signal-to-noise ratio ($>40$\,dB) and phase error ($<1^\circ$), as established by the robustness analysis in \cite{Zhong2025ExperimentallyProbingNonHermitian}.
Using the methods described above, we systematically construct lattices with various types of defects and present their corresponding spectra and eigenstates in the following subsections.

                 \subsection{Vacancy}~\label{subsec:vacancyexp}

Figure~\ref{fig:exp_vacancy} presents both the experimental observations and the corresponding tight-binding simulations of the energy eigenvalues and eigenstates for a single vacancy in a $7 \times 5$ acoustic lattice. In the acoustic lattice, the measured eigen-frequencies represent the energy spectrum, where the zero energy in the tight-binding model corresponds to the real part of the resonant frequency, $\Re(\omega_0) = 1038\,\mathrm{Hz}$. The experimental results are consistent with the predictions from the simulations. 
Although the absolute frequency separation between bands ($\sim 15$\,Hz) may appear small compared to the carrier frequency $\Re(\omega_0)$, the relevant energy scale governing localization is the gap relative to the effective hopping amplitudes. 
In our parameter regime, the ratio between the bandgap and hopping strength remains sufficiently large, resulting in a short decay length for the defect-bound states. 
Consequently, even in a finite lattice of size $7\times 5$, the defect modes remain strongly localized around the vacancy site, consistent with both numerical simulations and experimental observations.
As seen in Fig.~\ref{fig:exp_vacancy}, no in-gap bound states appear near the vacancy defect when $|m_0| > 2$, corresponding to a trivial insulator. By contrast, for $|m_0| < 2$, a pair of mid-gap bound states emerges, with their LDOS peaks localized near the defect site. These findings are also qualitatively consistent with the theoretical results with a single vacancy, displayed in Fig.~\ref{fig:2} (top row).

               \subsection{Schottky defect}~\label{subsec:schottkyexp}

Figure~\ref{fig:exp_Schottky} shows both the experimental observations and the corresponding tight-binding simulations of the energy eigenvalues and eigenstates for a single Schottky defect in a $6 \times 6$ acoustic square lattice. The experimental results are reasonably consistent with the ones from simulations. As seen in Fig.~\ref{fig:exp_Schottky}, when $|m_0| > 2$, two near-zero in-gap modes appear in the spectrum; however, their LDOS distributions show no localization near the defect, indicating a trivial insulating phase, insensitive to the presence of a Schottky defect. By contrast, for $|m_0| < 2$, a pair of mid-gap bound states emerges in the spectrum, with their LDOS peaks strongly localized around two sites, constituting the Schottky defect upon removing complementary orbitals from there. These findings are also in agreement with the ones from numerical simulation of the QWZ model with such a defect, see Fig.~\ref{fig:3} (top row).

            \subsection{Substitution}~\label{subsec:substitutionexp}

Figure~\ref{fig:exp_substitution} presents both the experimental observations and the corresponding tight-binding simulations of the energy eigenvalues and eigenstates for a single substitution defect in a $7 \times 5$ acoustic lattice. As in the theoretical analysis, we choose pairs of staggered mass parameters $(m_0^\mathrm{back}, m_0^\mathrm{sub})$ from two distinct parameter regimes of the QWZ model (see Secs.~\ref{sec:model} and~\ref{subsec:substitutiontheory} for details), resulting in a total of twelve possible configurations. The experimental results again show reasonable agreement with the simulations for all twelve configurations.

As shown in Fig.~\ref{fig:exp_substitution}, when the background mass parameter satisfies $|m_0^\mathrm{back}| > 2$, corresponding to a topologically trivial regime, no mid-gap bound states appear if $m_0^\mathrm{sub}$ also lies within the trivial regime (the LDOS shows no localization near the substitution site). In contrast, when $m_0^\mathrm{sub}$ is chosen from the topological regime, strongly localized bound states emerge near such point defect. On the other hand, when $|{m_0^\mathrm{back}}| < 2$, corresponding to a topological regime, localized mid-gap bound states near the substitution site are always observed regardless of the value of $m_0^\mathrm{sub}$. These findings are in qualitative agreement with previous theoretical analysis, reported in Fig.~\ref{fig:4}.

          \subsection{Interstitial}~\label{subsec:interstitialexp}

Figure~\ref{fig:exp_interstitial} shows both the experimental observations and the corresponding tight-binding simulations of the energy eigenvalues and eigenstates for a single interstitial defect in an otherwise $6 \times 6$ acoustic square lattice. As in the theoretical analysis, we choose pairs of staggered mass parameters $(m_0^\mathrm{back}, m_0^\mathrm{int})$ from two distinct parameter regimes of the QWZ model (see Sec.~\ref{subsec:interstitialtheory} for details), resulting in twelve possible configurations. For each such pair of mass parameters, the experimental results again show good agreement with the simulations.

As illustrated in Fig.~\ref{fig:exp_interstitial}, when the background parameter satisfies $|m_0^\mathrm{back}| > 2$, corresponding to a topologically trivial regime, highly localized mid-gap bound states with their LDOS peaked on the interstitial site appear only when $m_0^\mathrm{int}$ lies in the topologically nontrivial regime and has an opposite sign to $m_0^\mathrm{back}$, e.g., the ones for $m_0^\mathrm{back} = 2.5$ and $m_0^\mathrm{int} = -1.0$ shown in Fig.~\ref{fig:exp_interstitial} (i.b), which differs from the case of a substitution defect shown in Fig.~\ref{fig:exp_substitution}. By contrast, when $|m_0^\mathrm{back}| < 2$, corresponding to topological parameter regimes, localized mid-gap bound states with their LDOS peaked near the interstitial site are consistently observed regardless of the value of $m_0^\mathrm{int}$, similar to the behavior observed for the substitution defect. These findings are in qualitative agreement with theoretical predictions from Fig.~\ref{fig:5} (left column).

                \subsection{Frenkel pair}~\label{subsec:frenkelpairexp}

Figure~\ref{fig:exp_Frenkel} presents both the experimental observations and the corresponding tight-binding simulations of the energy eigenvalues and eigenstates for a single Frenkel pair defect, introduced near the center of a $7 \times 5$ acoustic lattice. As in the theoretical analysis, we choose pairs of staggered mass parameters $(m_0^\mathrm{back}, m_0^\mathrm{int})$ from two distinct parameter regimes of the QWZ model (see Secs.~\ref{sec:model} and~\ref{subsec:frenkeltheory}), yielding a total of twelve possible configurations. The experimental results show good agreement with the simulations for each such configuration.

As shown in Fig.~\ref{fig:exp_Frenkel}, when the background parameter satisfies $|m_0^\mathrm{back}| > 2$, corresponding to a topologically trivial regime, no mid-gap bound states appear if $m_0^\mathrm{int}$ also lies within the trivial regime. In contrast, when $m_0^\mathrm{int}$ is chosen from the topological regime, strongly localized bound states emerge near such defect. However, we realize that such modes, observed when $m_0^\mathrm{int}$ and $m_0^\mathrm{back}$ are of same sign with the former (latter) one falling in the topological (trivial) parameter regime are solely due to the smallness of the system, which are absent in larger systems, as found in our numerical simulations. On the other hand, when $|m_0^\mathrm{back}| < 2$, corresponding to a topological regime, localized mid-gap bound states are always observed regardless of the value of $m_0^\mathrm{int}$. Otherwise, these findings are in qualitative agreement with theoretical predictions, shown in Fig.~\ref{fig:5} (right column). Finally, notice that the LDOS associated with such mid-gap modes is distributed between the constituting vacancy and interstitial sites (highly localized only around the constituting interstitial site) when $m_0^\mathrm{back}$ falls within the topological (trivial) regime, as observed in Fig.~\ref{fig:exp_Frenkel}, which was also predicted theoretically in Sec.~\ref{subsec:frenkeltheory}.

%%%%%%%%%%%%%%%%%%%%%%%%%%%%%%%%%%%%%%%%%%%%%%%%%%%%%%%%%%%%%%%%%%%%%%
%%%%%%%%%%%%%%%%%%%%%%%%%%%%%%%%%%%%%%%%%%%%%%%%%%%%%%%%%%%%%%%%%%%%%%
%%%%%%%%%%%%%%%%%%%%%%%%%%%%%%%%%%%%%%%%%%%%%%%%%%%%%%%%%%%%%%%%%%%%%%
\begin{figure*}[t!]
    \centering
    \includegraphics[width=1.00\linewidth]{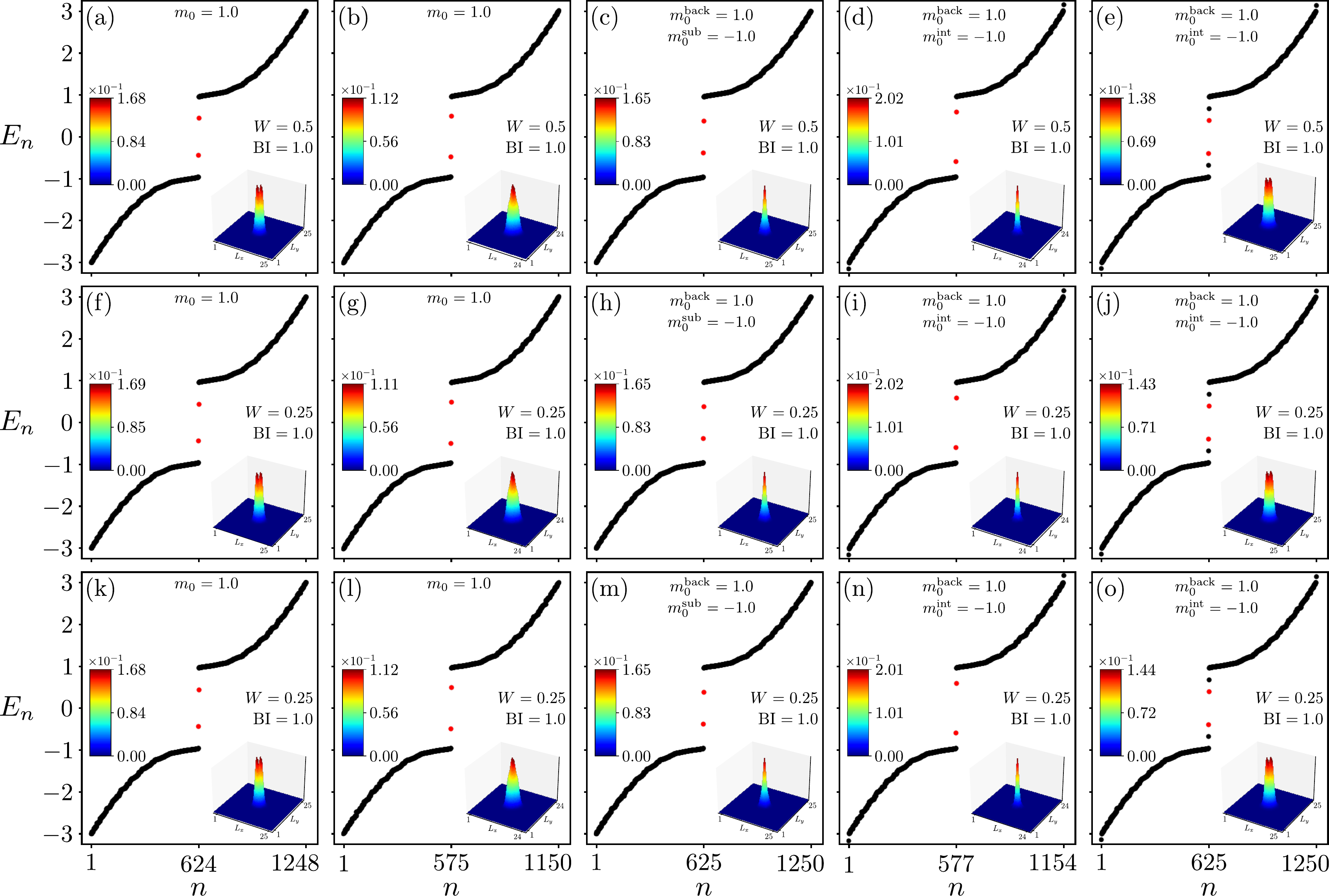}
    \caption{
             Energy eigenvalues ($E_n$) as a function of its index ($n$) with (a) a vacancy, (b) a Schottky defect, (c) a substitution, (d) an interstitial, and (e) a Frenkel pair with the same parameter values and system sizes as in Figs.~\ref{fig:2}(b), \ref{fig:3}(f), \ref{fig:4}(i.e), \ref{fig:5}(i.e), and~\ref{fig:5}(ii.e), respectively, in the presence of random point-like charge impurities, characterized by their strength $W=0.5$ (1/4th of the band gap in the pristine square lattice). The eigenspectrum and the local density of states for two particle-hole symmetric closest-to-zero-energy modes (shown in red) are presented for a specific disorder realization. On the other hand, the Bott index (BI) is obtained after averaging over 100 random and independent disorder realizations. Analogous results are shown in panels (f)-(j) for random mass disorder and in panels (k)-(o) for random bond disorder, where $W$ denotes the strength of the corresponding disorder. Altogether these results establish robustness of all types of defect-bound midgap modes against sufficiently weak generic disorder.
             For discussion on the results consult Appendix~\ref{appensec:disorder}.           
    }
    \label{fig:append}
\end{figure*}
%%%%%%%%%%%%%%%%%%%%%%%%%%%%%%%%%%%%%%%%%%%%%%%%%%%%%%%%%%%%%%%%%%%%%%
%%%%%%%%%%%%%%%%%%%%%%%%%%%%%%%%%%%%%%%%%%%%%%%%%%%%%%%%%%%%%%%%%%%%%%
%%%%%%%%%%%%%%%%%%%%%%%%%%%%%%%%%%%%%%%%%%%%%%%%%%%%%%%%%%%%%%%%%%%%%%

\section{Summary and discussion}~\label{sec:summary}

To summarize, through comprehensive theoretical and experimental investigations, we have established  ordinary, but ubiquitous lattice defects as probes of topological crystals as well as any change in local topological environment in normal and topological insulators in terms of mid-gap modes bound to such lattice defects that are well separated from the remainder of the bulk states and are immune to weak random charge impurities. Sharp localization of topological mid-gap modes in the close proximity to ordinary lattice defects is further established from their inverse participation ratio (IPR) in Appendix~\ref{appensec:IPR}. We establish these generic outcomes from the existence of robust mid-gap bound states near the ordinary lattice defects like vacancy, Schottky defect, substitution, interstitial, and Frenkel pair. Although for the sake of demonstration, here we exclusively considered insulating systems that are devoid of the time-reversal symmetry, it does not play any role in any conclusion of ours. Therefore, our findings should be equally pertinent in any lattice from any symmetry class and of arbitrary physical dimension~\cite{AZ:1, AZ:2, AZ:3, AZ:4, AZ:5, AZ:6}, which can especially be of particular interest for trapping localized Majorana modes around such ordinary lattice defects in higher-dimensional topological superconductors. Notice that with suitable redefinition of the spinor basis, the QWZ model also describes the lattice-regularized topological $p+ip$ pairing when there exists an underlying Fermi surface around the $\Gamma$ (for $-2<m_0/t_0<0$) or the ${\rm M}$ (for $0<m_0/t_0<2$) point of the Brillouin zone~\cite{model:4}. It should be noted that in the presence of any ordinary lattice defect, even number of particle-hole symmetric bound states appear in pair. Consequently, existence of such defect-bound localized modes does not led to charge fractionalization, which necessarily requires the presence of unpaired (odd number of) localized mode~\cite{jackiw-rebbi}.

These findings open vastly unexplored avenues for both fundamental and applied research. For example, in topological superconductors, by tuning the spatial localization and separation of such ordinary-defect-bound Majorana modes, one can also accomplish efficient and practical routes for their braiding, an essential ingredient for topological quantum computations~\cite{braiding:1, braiding:2}, particularly on optical lattices and in designer quantum materials, since topological models have already been emulated in former systems~\cite{OL:1, OL:2}. Moreover, the concept of such mid-gap bound states can be naturally extended to dynamic or Floquet crystals as well as non-Hermitian lattices, with the possibility of observing ordinary defect-localized non-Hermitian skin effects in the latter systems, similar to the ones recently observed near dislocation lattice defects~\cite{dislocation:19, dislocation:20}. We reserve these promising directions for future investigations. Our experimental observation of defect-bound modes on acoustic lattices opens unexplored avenues for ordinary defect-engineered topological devices, which can be extended to the territory of other classical meta-materials, such as topolectric circuits~\cite{topolectric:1, topolectric:2, topolectric:3}, and photonic~\cite{photonic:1, photonic:2, photonic:3} and phononic~\cite{phononic:1, phononic:2, phononic:3} lattices,  and possibly on photonic~\cite{photonicring:1} and plasmonic~\cite{plasmonicring:1} rings to emulate ordinary lattice defect-based topological phenomena.

\acknowledgments

A.M.\ and B.R.\ were supported by NSF CAREER Grant No.\ DMR-2238679 of B.R. Y.J.\ thanks the support of startup funds from Pennsylvania State University and NSF CMMI awards 2039463, 195122, and 2401236. B.R.\ thanks Vladimir Juri\v ci\' c and Christopher A.\ Leong for critical readings of the manuscript.

\section*{Data availability}

All the numerical codes and data used in this work are available on Zenodo~\cite{datacode:aiden}.

\appendix

\section{Stability of mid-gap defect modes against disorder}~\label{appensec:disorder}

In this Appendix, we address the stability of mid-gap modes bound to ordinary lattice defects, studied in this work, against disorder. For the purpose of demonstration, we consider one set of parameter values for each such lattice defect for which there exists robust mid-gap bound states around the corresponding defect core in clean systems. Next we add on-site random charge impurities (the dominant source of the elastic scattering in any real material) given by the Hamiltonian 
\begin{equation}~\label{eq:disorder}
H_{\rm dis}=\sum_{\vec{r}_j} V(\vec{r}_j) \; c^\dagger_j \tau_0 c_j. 
\end{equation}
At each lattice site (indexed by $j$), located at ${\vec r}_j$, we draw $V(\vec{r}_j)$ randomly and independently from a uniform box distribution $[-W/2,W/2]$, where the box width $W$ measures the strength of disorder. In a sufficiently large system ${\rm tr}(V(\vec{r}_j))=0$, where the `tr' denotes a trace over all the sites, ensuring the particle-hole symmetry in the thermodynamic limit. However, in smaller systems (as is the case in our numerical simulations), such a condition does not necessarily hold for every disorder configuration. In order to ensure the particle-hole symmetry even in smaller systems that makes our computation of the Bott index and identification of mid-gap states easier, we define a quantity $\delta={\rm tr}(V(\vec{r}_j))/L^2$ and replace $V(\vec{r}_j)$ by $V(\vec{r}_j)-\delta \times {\boldsymbol I}_L$, where ${\boldsymbol I}_L$ is an $L \times L$ identity matrix. Such a slightly modified disorder potential maintains the particle-hole symmetry in every system for any value of $W$. With such a modified disorder potential, we add $H_{\rm dis}$ from Eq.~\eqref{eq:disorder} to either $H^{\rm SL}_{\rm TB}$ from Eq.~\eqref{eq:realspaceTB} or $H^{\rm SL, gen}_{\rm TB}$ from Eq.~\eqref{eq:Method1}, depending on the nature of the lattice defect and repeat the numerical exact diagonalization procedure. For all the cases, we choose $W$ to be $1/4$th of the band gap in a pristine (defect-free and clean) square lattice with the same set of parameter values. We showcase only a few cases in Figs.~\ref{fig:append}(a)-\ref{fig:append}(d) that nonetheless convey an important message that the mid-gap modes bound to ordinary lattice defects are robust against at least sufficiently weak disorder in the system.

%%%%%%%%%%%%%%%%%%%%%%%%%%%%%%%%%%%%%%%%%%%%%%%%%%%%%%%%%%%%%%%%%%%%%%
%%%%%%%%%%%%%%%%%%%%%%%%%%%%%%%%%%%%%%%%%%%%%%%%%%%%%%%%%%%%%%%%%%%%%%
%%%%%%%%%%%%%%%%%%%%%%%%%%%%%%%%%%%%%%%%%%%%%%%%%%%%%%%%%%%%%%%%%%%%%%
\begin{figure*}[t!]
    \centering
    \includegraphics[width=1.00\linewidth]{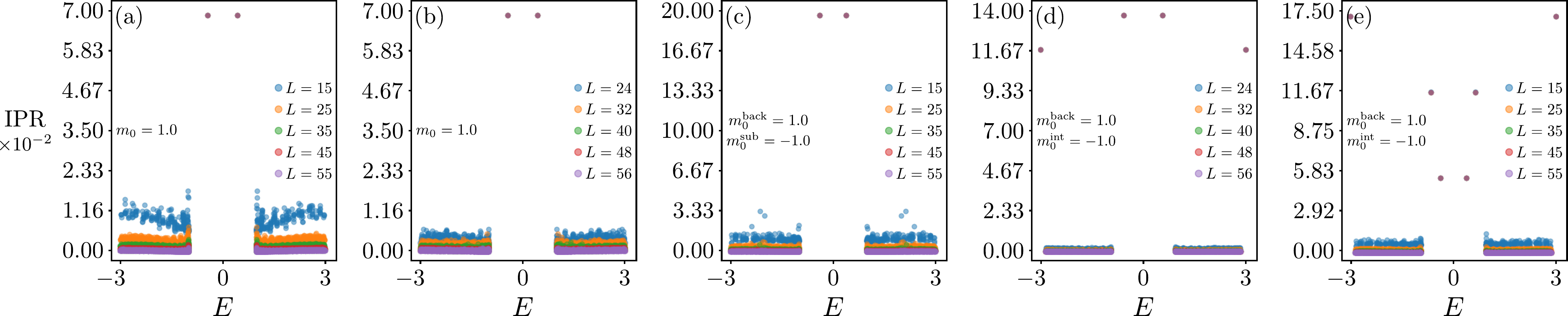}
    \caption{Inverse participation ratio (IPR) spectra over the entire range of energy ($E$) in the presence of (a) a vacancy, (b) a Schottky defect, (c) a substitution, (d) an interstitial,
    and (e) a Frenkel pair for the same parameter values in Figs.~\ref{fig:2}(b), \ref{fig:3}(f), \ref{fig:4}(i.e), \ref{fig:5}(i.e), and~\ref{fig:5}(ii.e), respectively, with varying system size $L$ (linear dimension of the square lattice in each direction). Notice that the IPR for the mig-gap states in each case is insensitive to the system size $L$, confirming their localized nature around the defect core. For discussion on the results consult Appendix~\ref{appensec:IPR}.           
    }
    \label{fig:IPR}
\end{figure*}
%%%%%%%%%%%%%%%%%%%%%%%%%%%%%%%%%%%%%%%%%%%%%%%%%%%%%%%%%%%%%%%%%%%%%%
%%%%%%%%%%%%%%%%%%%%%%%%%%%%%%%%%%%%%%%%%%%%%%%%%%%%%%%%%%%%%%%%%%%%%%
%%%%%%%%%%%%%%%%%%%%%%%%%%%%%%%%%%%%%%%%%%%%%%%%%%%%%%%%%%%%%%%%%%%%%%

It should be noted that the stability of defect-bound midgap modes arises from their spectral gap with rest of the bulk states. Therefore, we expect such modes to display robustness against sufficiently weak generic disorder. To establish this claim, next we scrutinize the stability of defect-bound miggap modes against random mass disorder, which can be implemented following the scheme discussed above with $\tau_0 \to \tau_3$ in Eq.~\eqref{eq:disorder}. The strength of random mass disorder $W$ is chosen to be $1/4$th of $m_0$ in the presence of vacancy and Schottky defects, and $1/4$th of $m^{\rm back}_0$ with underlying substitution, interstitial, and Frenkel pair defects. The results are shown in Figs.~\ref{fig:append}(e)-\ref{fig:append}(h), demonstrating the robustness of such defect-bound midgap modes against sufficiently weak random mass disorder. In addition, we also consider random bond disorder for which the nearest-neighbor uniform hopping amplitude $t$ between the orbitals with opposite parities appearing in Eqs.~\eqref{eq:QWZBloch} and~\eqref{eq:dvec} is accompanied by a random hopping amplitude chosen from a uniform box distribution $[-W/2, W /2]$ with $W=t/4$. The results are shown in Figs.~\ref{fig:append}(i)-\ref{fig:append}(l), showing that all types of defect-bound miggap modes are stable against sufficiently random bond or hopping disorder.

\section{Inverse participation ratio}~\label{appensec:IPR}

In this Appendix, we demonstrate the localization of mid-gap states in the close vicinity of the ordinary lattice defect centers from their IPR. For any eigenstate at energy $E$, denoted by $\Psi_E(\vec{r})$, the corresponding IPR is defined as 
\begin{equation}
{\rm IPR}= \sum_{\vec{r}} \left( \sum_{\kappa} \bigg| \Psi^{\kappa}_E(\vec{r}) \bigg|^2 \right)^2,
\end{equation}
where $\Psi^{\kappa}_E(\vec{r})$ is the amplitude of a given eigenstate with the internal degrees of freedom $\kappa$ at position $\vec{r}$. For the QWZ model, $\kappa=1$ and $2$ correspond to two orbitals with opposite parities. We compute the IPR for all the states at various energies $E$ as a function of $L$, where $L$ is the linear dimension of the square lattice with periodic boundary condition in each direction. Physically, for the states localized around a few sites, the IPR is almost independent of the system size. On the other hand, for extended states that uniformly spread uniformly over the entire system the IPR scales as $1/N$, where $N$ is the total number of sites in the system~\cite{salib-das-roy}. The results are shown in Fig.~\ref{fig:IPR}, confirming that the mid-gap modes are localized as observed from their scaling of IPR with varying system size, which reside around the defect cores that we have previously observed from their local density of states.

\end{document}